\documentclass[usenatbib]{mn2e}   
\pdfoutput=1 
\usepackage[pdftex]{graphicx} 
\usepackage{times}
\usepackage{rotate} 
\usepackage{amssymb} 
\usepackage{rotating}
\usepackage{epstopdf} 

\title[Black hole accretion modes since $z \sim 1$]{Investigating evidence for
  different black hole accretion modes since redshift $\mathbf{z\sim1}$}
\author[Georgakakis et al.]   {A. Georgakakis$^{1, 2}$,
  P. G. P\'erez-Gonz\'alez$^{3, 4}$,  N. Fanidakis$^5$, 
  M. Salvato$^1$, J. Aird$^6$,
\newauthor H. Messias$^7$, J. M. Lotz$^8$, G. Barro$^9$, Li-Ting
Hsu$^1$, K.  Nandra$^1$, D. Rosario$^1$, \newauthor M. C. Cooper$^{10}$, 
 D. D. Kocevski$^{11}$, J. A. Newman$^{12}$
\\ \\ 
 $^1$Department of Physics and Astronomy
 Planck  Institut f\"{u}r Extraterrestrische Physik, Giessenbachstra\ss e, 85748 Garching, Germany\\ 
 $^2$IAASARS, National Observatory of Athens, GR-15236 Penteli, Greece\\ 
 $^3$Departamento de Astrof\'{\i}sica, Facultad de CC. F\'{\i}sicas, Universidad Complutense de Madrid, E-28040 Madrid, Spain\\
 $^4$Steward Observatory, The University of Arizona, 933 N. Cherry Ave., Tucson, AZ 85721, USA\\
 $^5$Max Planck Institut f\"{u}r Astronomie , K\"{o}nigstuhl 17,  D-69117, Heidelberg, Germany\\
 $^6$Extragalactic \& Cosmology Group, Rochester Building, Department of Physics, University of Durham, Science Laboratoriesm South Roadm Durham DH1 3LE.\\
 $^7$Departamento de Astronom\'ia, Av. Esteban Iturra 6to piso, Facultad de Ciencias F\'isicas y Matem\'aticas, Universidad de Concepci\'on, Chile.\\
 $^8$Space Telescope Science Institute, 3700 San Martin Drive, Baltimore, MD 21218, USA\\ 
 $^9$UCO/Lick Observatory, University of California, Santa Cruz, 1156 High Street, Santa Cruz,  CA 95064\\
 $^{10}$Center for Galaxy Evolution, Department of Physics and Astronomy,
University of California, Irvine, 4129 Frederick Reines Hall Irvine, CA 92697 USA\\
$^{11}$University of Kentucky, Department of Physics and Astronomy, 177 Chemistry-Physics Building, Lexington, KY 40506-0055, USA\\
 $^{12}$Department of Physics and Astronomy \& Pittsburgh Particle
Physics, Astrophysics and Cosmology Center (PITT PACC), University of Pittsburgh, \\Pittsburgh, PA 15260, USA
}

\begin{document}
\maketitle

\begin{abstract}  Chandra  data  in  the COSMOS,  AEGIS-XD  and  4\,Ms
Chandra Deep Field South  are combined with multiwavelength photometry
available in those  fields to determine the rest-frame  $U-V$ vs $V-J$
colours of X-ray AGN hosts in the redshift intervals $0.1<z<0.6$ (mean
$\overline{z}=0.40$) and $0.6<z<1.2$ (mean $\overline{z}=0.85$).  This
combination of colours provides an effective and least model-dependent
means  of  separating  quiescent  from  star-forming,  including  dust
reddened, galaxies.   Morphological information emphasises differences
between AGN populations split by  their $U-V$ vs $V-J$ colours. AGN in
quiescent  galaxies  consist   almost  exclusively  of  bulges,  while
star-forming  hosts  are equally  split  between  early and  late-type
hosts.  The  position of AGN  hosts on the  $U-V$ vs $V-J$  diagram is
then  used to  set limits  on the  accretion density  of  the Universe
associated with  evolved and star-forming systems  independent of dust
induced biases.   It is found  that most of  the black hole  growth at
$z\approx0.40$  and  0.85   is  associated  with  star-forming  hosts.
Nevertheless,  a  non-negligible  fraction  of  the  X-ray  luminosity
density,  about  15-20\%, at  both  $\overline{z}=0.40$  and 0.85,  is
taking place in galaxies in the quiescent region of the $U-V$ vs $V-J$
diagram.  For  the low-redshift  subsample, $0.1<z<0.6$, we  also find
tentative evidence,  significant at the  $2\sigma$ level, that AGN  split by
their  $U-V$   and  $V-J$  colours  have   different  Eddington  ratio
distributions.  AGN in blue  star-forming hosts dominate at relatively
high Eddington ratios.  In contrast, AGN in red quiescent hosts become
increasingly important  as a fraction  of the total  population toward
low Eddington  ratios.  At higher redshift,  $z>0.6$, such differences
are significant at the $2\sigma$ level only for sources with Eddington
ratios $\ga10^{-3}$.   These findings are consistent  with scenarios in
which  diverse accretion  modes are  responsible for  the  build-up of
supermassive black holes at the centres of galaxies.  We compare these
results with the predictions  of the {\sc galform} semi-analytic model
for  the  cosmological evolution  of  AGN  and  galaxies.  This  model
postulates  two black  hole fuelling  modes,  the first  is linked  to
star-formation events and the  second takes place in passive galaxies.
{\sc galform} predicts  that a substantial fraction of  the black hole
growth  at $z<1$ is  associated with  quiescent galaxies,  in apparent
conflict with the observations.  Relaxing the strong assumption of the
model that passive AGN hosts have zero star-formation rate could bring
those predictions in better agreement with the data.
\end{abstract}

\section{Introduction}\label{sec_intro}

In recent years observational data established that supermassive black
holes   (SMBHs)    are   nearly   ubiquitous    in   local   spheroids
\citep[e.g.][]{Magorrian1998}.  Moreover, correlations were discovered
between the masses  of those black holes and  the stellar component of
the  bulges  in  which  they  reside  \citep[e.g.][and references therein]{Kormendy_Ho2013}.   These
empirical correlations have been combined with large galaxy surveys to
place tight constraints  on the mass function of  dormant SMBHs in the
nearby   Universe  \citep[e.g.][]{Kelly_Merloni2012}.    What  remains
unclear however, is how the  relic SMBHs we observe in nearby galaxies
grow  their  mass  across  cosmic  time.  One  way  to  approach  this
question is to  conduct population studies of the  galaxies that host
active   SMBHs   at  different   redshifts.    The  properties   (e.g.
morphology,  environment)  of the  galaxies  with  an Active  Galactic
Nucleus (AGN) provide information  on the physical conditions on large
(kpc to Mpc) scales which may be relevant to the fuelling of the SMBH.

Morphological studies  for example, find  that X-ray AGN hosts  in the
redshift range $z\approx0.5-2$ have diverse morphologies (spiral,
elliptical, disturbed) with a relative mix that is similar to that of
mass-matched  non-AGN  galaxy samples  \citep[e.g.][]{Georgakakis2009,
Cisternas2011, Kocevski2012}.  This  suggests that major mergers, 
which  are expected  to be  associated with  morphologically disturbed
systems, cannot  be the only channel  for growing black  holes at the
centres of  galaxies.  Other  mechanisms, e.g.  minor  interactions or
secular processes,  must also contribute  to the accretion  density of
the  Universe.   This conclusion  is  also  supported  by large  scale
structure  studies, which estimate  mean dark  matter halo  masses for
X-ray AGN  in the range $\rm  \log (M/M_\odot)\approx12.5-13.5$.  This
mass  interval is  larger than  expected  if black  hole accretion  is
triggered   by   major   mergers   only   \citep[e.g.][]{Allevato2011,
Mountrichas2012,      Mountrichas2013}.       Recent      work      by
\cite{Fanidakis2013} indeed shows that  the clustering of X-ray AGN at
$z\la1.5$  is consistent  with two  channels for  growing  SMBHs.  The
first is associated with star-formation  events in the host galaxy and
the second is related to  quiescent galaxies in massive halos.  In the
modelling of  \cite{Fanidakis2013} star-formation  is a proxy  to cold
gas availability.   Galaxies with abundant cold gas  supplies can form
stars and grow their central black holes at a high rate.  In contrast,
the SMBHs  of evolved galaxies  that are devoid  of cold gas  can only
grow  slowly  via  the  accretion  of  hot  gas  from  a  quasi-static
atmosphere.   Evidence   for  a   dichotomy  in  the   accretion  rate
distribution  of  narrow  optical   emission-line  AGN  based  on  the
star-formation  history of their  hosts is  reported at  low redshifts
\citep{Kauffmann_Heckman2009}.   This finding further  supports claims
for diverse  AGN fuelling modes  and suggests that  the star-formation
properties  of AGN hosts  hold important  information on  the physical
conditions under which black holes at the centres of galaxies build-up
their mass.

The   evidence  above   has  motivated   efforts  to   understand  the
star-formation level  of AGN hosts  at higher redshift to  explore how
black  holes are  fuelled as  a function  of cosmic  time.  Population
studies have established that the build-up of black holes and galaxies
are related in  a statistical sense when integrated  in a cosmological
volume.  The star-formation rate density \citep{Hopkins_A2006} and the
accretion  luminosity  density  \citep{Aird2010} follow  very  similar
evolution  patterns with redshift.   There is  also evidence  that the
cosmological  evolution of  the AGN  space density  is related  to the
increase  with redshift  of the  average specific  star-formation rate
(star-formation   rate   per    unit   stellar   mass)   of   galaxies
\cite[e.g.][]{Georgakakis2011}.     Far-IR/sub-mm   observations   with
Herschel   extended  measurements  of   the  star-formation   rate  of
individual AGN to high redshift and also bright accretion luminosities
(e.g. luminous  QSOs) where  other indicators (e.g.   optical spectral
features, broad-band colours) become  unreliable. Although it does not
appear that there is a  one-to-one correspondence between the level of
star-formation   and    the   accretion   power   \citep{Mullaney2012,
  Rosario2012}, X-ray  AGN are on average associated  with galaxies on
the  main  star-formation  sequence \citep{Santini2012,  Mullaney2012,
  Rovilos2012, Rosario2013}.  

In  a   typical  Herschel   far-IR/sub-mm  survey  field   however,  a
substantial  fraction of  the  AGN population  lies  below the  formal
detection limit.   Stacking the far-IR/sub-mm counts  at the positions
of X-ray sources  is used extensively to reach  deeper flux limits and
explore  the  star-formation  properties  of  the  entire  population.
Although valuable, this approach  has the limitation that it estimates
only the mean far-IR/sub-mm properties  of AGN hosts and provides only
limited   information  on  the   underlying  distribution.    A  small
sub-population of AGN not associated with high specific star-formation
rate events  is likely  to be averaged  out in  far-IR/sub-mm stacking
studies.  Optical observations for example, show that a large fraction
of  the X-ray AGN  hosts at  $z\la1$ lie  in the  red sequence  of the
colour-magnitude  diagram \citep[e.g.][]{Hickox2009, Georgakakis2011},
which includes  a large fraction  of passive galaxies.   Although dust
can redden the  broad-band colours of galaxies, it  cannot account for
the entire population  of AGN hosts on the red  sequence of the colour
magnitude  diagram.    \cite{Cardamone2010cmd}  for example,  find
evidence for  a bi-modal $U-V$ rest-frame colour  distribution for AGN
hosts at  $z\approx1$, once  the impact of  dust is accounted  for via
fitting  templates  to  the  observed spectral  energy  distributions.
\cite{Mignoli2004}  argue that  obscured X-ray  selected QSO  hosts at
$z\approx1-2$ have  rest-frame optical light profiles  that follow the
de vaucouleurs  law.  This  is interepreted as  evidence that  a large
fraction of the obscured QSO  population at $z\approx1-2$ is hosted by
bulge-dominated  galaxies, possibly  quiescent  ellipticals \citep[but
see ][]{Hutchings2002}.

In this paper we place limits on the fraction of the accretion density
of the Universe associated with quiescent, low specific star-formation
galaxies  in the  redshift range  0.1--1.2.   X-ray data  are used  to
select AGN and their rest-frame  broad-band colours are adopted as the
least  model-dependent  method  to  discriminate between  evolved  and
actively star-forming  hosts.  Dust reddening issues  are mitigated by
placing  X-ray AGN  hosts on  the $U-V$  vs $V-J$  (UVJ) colour-colour
diagram \citep{Williams2009, Patel2012}.   This combination of colours
is  least sensitive  to  dust extinction  and  has been  shown to  be
effective in  separating early-type, low-specific  star-formation rate
galaxies from  actively star-forming, including  dust-reddened systems
\citep{Williams2009}. Throughout  this paper we adopt $\rm  H_{0} = 70
\,  km  \,  s^{-1} \,  Mpc^{-1}$,  $\rm  \Omega_{M}  = 0.3$  and  $\rm
\Omega_{\Lambda} = 0.7$.

\section{X-ray AGN sample}\label{sec_xagn}

We combine  data from X-ray surveys with  different characteristics in
terms of  area coverage and X-ray  depth. These are  the 4\,Ms Chandra
Deep Field South \citep[CDFS;][]{Xue2011}, the Chandra 800\,ks survey of
the AEGIS  field (AEGIS-XD;  Nandra et al.   in prep) and  the Chandra
survey  of  the  COSMOS field  \citep[C-COSMOS,][]{Elvis2009}.   These
samples provide  sufficient coverage of  the $L_X-z$ plane  to explore
the evolution of the properties of X-ray AGN hosts since $z\approx1$.

\begin{figure}
\begin{center}
\includegraphics[height=0.9\columnwidth]{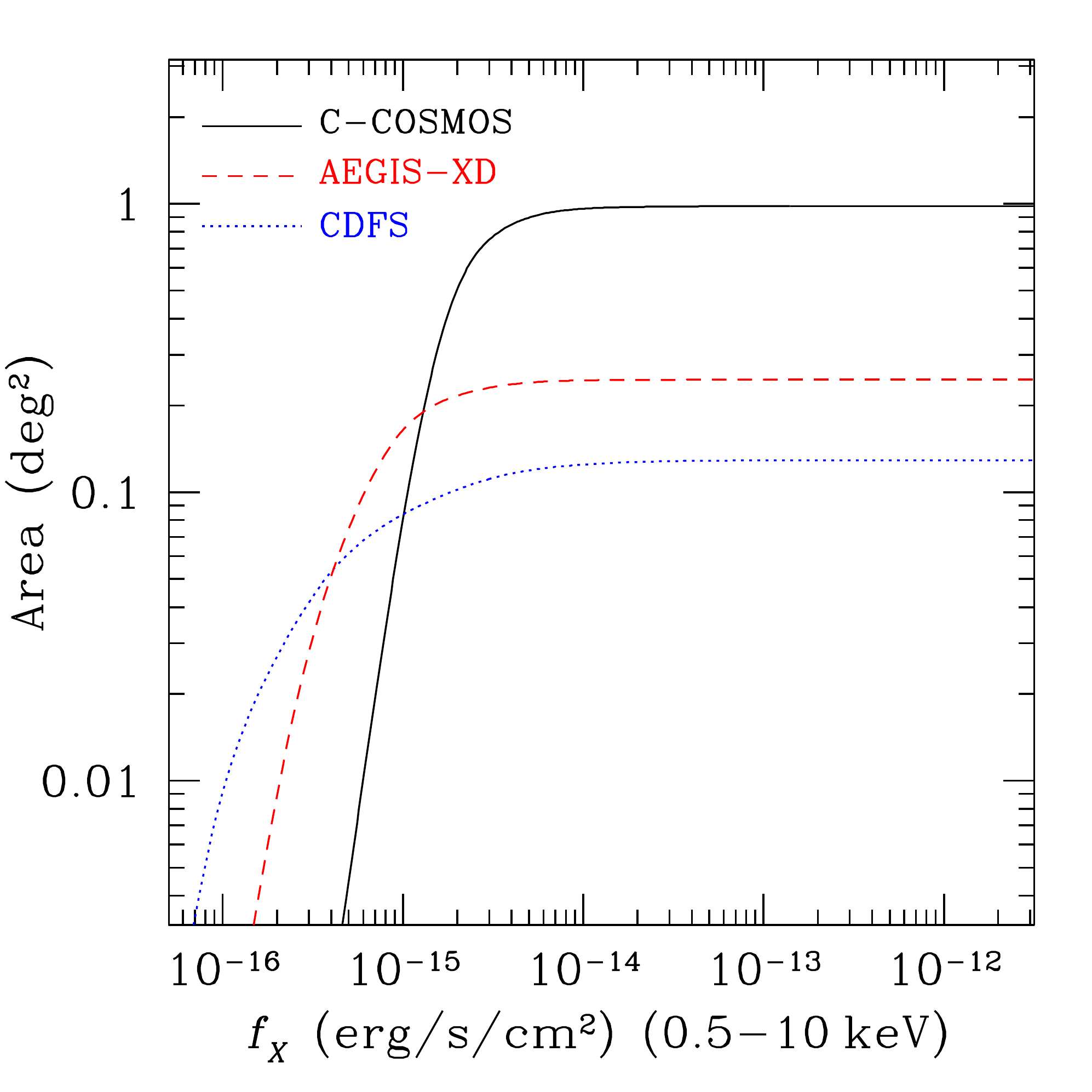}
\end{center}
\caption{Sensitivity  curves in  the 0.5-10\,keV  energy band  for the
  CDFS, AEGIS-XD and C-COSMOS fields.}\label{fig_curve}
\end{figure}

The  Chandra observations  of  the CDFS,  AEGIS-XD  and C-COSMOS  were
analysed in  a homogeneous  way by applying  the reduction  and source
detection methodology  described by \cite{Laird2009}.   The motivation
for  this  is to  have  a  homogeneous  and well  characterised  X-ray
selection  function accross  the three  fields, which  is advantageous
when  studying  the  statistical  properties  of  the  X-ray  detected
population.  A total  of 569, 937 and 1584  X-ray sources are detected
in the CDFS,  AEGIS-XD and C-COSMOS, respectively, in  at least one of
the   soft  (0.5-2\,keV),  hard   (2-7\,keV),  full   (0.5-7\,keV)  or
ultra-hard (5-7\,keV)  spectral bands  to the Poisson  false detection
threshold of  $4\times10^{-6}$ (see Laird  et al.  2009  for details).
The number  of X-ray  detections in the  CDFS and C-COSMOS  is smaller
than that in the catalogues published  by Xue et al.  (2011) and Elvis
et al. (2009),  respectively.  This is because of  the lower detection
threshold adopted in those  studies. The optical identification of the
X-ray   sources   was   based   on   the   Likelihood   Ratio   method
\citep{Sutherland_and_Saunders1992}.   The  CDFS  X-ray  sources  were
cross-matched   with   the   MUSYC   optical   photometric   catalogue
\citep{Cardamone2010}.   In  the case  of  the  AEGIS-XD  we used  the
IRAC-$\rm  3.6\mu  m$  selected multi-waveband  photometric  catalogue
provided    by    the    Rainbow   Cosmological    Surveys    Database
\citep{Perez2008,  Barro2011a,  Barro2011b}.   The  identification  of
C-COSMOS X-ray  sources used the  $I$-band selected optical  sample of
\cite{Capak2007}   and   the   IRAC-$\rm   3.6\mu  m$   catalogue   of
\cite{Sanders2007}.

Extensive spectroscopic campaigns have  been carried out in the fields
of choice.  In  the CDFS we used the  spectroscopic redshifts compiled
by \cite{Cardamone2010} as part of the MUSYC multiwavelength catalogue
release. Spectroscopic  redshift measurements of X-ray  sources in the
AEGIS field are primarily  from the DEEP2 \citep{Newman2012} and DEEP3
galaxy redshift surveys \citep{Cooper2011_deep3_1, Cooper2012_deep3_2}
as well  as observations  carried out at  the MMT using  the Hectospec
fibre  spectrograph \citep{Coil2009}.  Redshifts  in the  C-COSMOS are
from  the   public  releases  of  the   VIMOS/zCOSMOS  bright  project
\citep{Lilly2009}   and  the   Magellan/IMACS   observation  campaigns
\citep{Trump2009}, as  well as the compilation of  redshifts for X-ray
sources presented by \cite{Brusa2010}.

We select X-ray sources with ${\cal R}<24$\,mag, where $\cal R$ stands
for  either  the  MUSYC  $R$-band   in  the  case  of  the  CDFS,  the
$r^{\prime}$  filter  of  the  CFHT  (Canada-France-Hawaii  Telescope)
Megaprime camera for the AEGIS-XD or the Subaru Suprime-Cam instrument
$r^+$ band  in the  case of C-COSMOS.   At these magnitude  limits the
spectroscopic identification rate of  the CDFS, AEGIS-XD, and C-COSMOS
X-ray sources is 78 (176/224), 70 (288/414) and 75 (726/962) per cent,
respectively (see  Table \ref{tab_sample}).  We also  limit the sample
to X-ray sources with spectroscopic redshift measurements in the range
$0.1-1.2$  (see Table  \ref{tab_sample} for  the total  number of
sources).  X-ray  sources brighter  than ${\cal R}  =24$\,mag without
spectroscopic redshift  measurements are  used only indirectly  in the
analysis.  The photometric redshift probability distributions (PDZ) of
those sources are integrated to estimate corrections for spectroscopic
incompleteness in  the calculation  of the space  density of  AGN (see
section \ref{sec_xlf}).  The X-ray  AGN photometric redshifts and PDZs
are from \cite{Salvato2011} for C-COSMOS  and Nandra et al.  (in prep)
for  the AEGIS-XD.   The methodology  described in  those publications
have also been  applied to the MUSYC photometry  to determine PDZs for
the  CDFS X-ray  sources.  A  by-product of  the  photometric redshift
determination   is  the  characterisation   of  the   Spectral  Energy
Distribution  (SED) of  X-ray AGN,  e.g.  host  galaxy type,  level of
optical  extinction,  level  of  the  AGN component  relative  to  the
underlying  host galaxy.   The  latter information  is  used in  later
sections  to  identify sources  for  which  the  AGN radiation  likely
contaminates the host galaxy light.

The CDFS,  AEGIS-XD and C-COSMOS  spectroscopic X-ray AGN  samples are
split into two redshift bins, 0.1--0.6 and 0.6--1.2, with medians 0.40
and 0.85, respectively.  We choose to select sources in the 0.5-7\,keV
spectral  band for  both redshift  sub-samples.  The  total  number of
X-ray sources  in each field  is shown in Table  \ref{tab_sample}. The
X-ray sensitivity curves are estimated by extrapolating the background
counts and exposure  maps in the 0.5-7\,keV band  to the limiting flux
of  a source  in the  0.5-10\,keV energy  range.  The  resulting X-ray
sensitivity  curves are plotted  in Figure  \ref{fig_curve} for  the 3
survey fields used in the analysis.

High resolution  imaging observations from the  Hubble Space Telescope
(HST) are also used to explore  the morphology of the host galaxies of
X-ray sources.  The  Advanced Camera for Surveys (ACS)  aboard HST has
surveyed  the  central  most  sensitive  part  of  the  CDFS  in  four
passbands, F435W, F606W, F775W and F850LP, with corresponding exposure
times 7200, 5450, 7028  and 18200\,s, respectively.  The survey setup,
data    reduction    and   source    detection    is   described    by
\cite{Giavalisco2004}.    The  estimated  $10\,\sigma$   point  source
limiting magnitude in the F775W filter is about 27\,mag. About 75\% of
the 4Ms  CDFS X-ray sources overlap  with the HST  survey region.  The
AEGIS-XD field  also has  HST/ACS imaging in  the F606W  (2260\,s) and
F814W  (2100\,s)  filters  \citep{Lotz2008_aegis}. These  observations
cover  a subregion of  the AEGIS-XD  that includes  about 65\%  of the
X-ray sources. The 5\,sigma limiting magnitudes for a point source are
$V_{F606W}= 28.14$  (AB) and $I_{F814W}  = 27.52$ (AB).  The  HST 
surveyed  the  COSMOS   field  with  the  ACS  in   the  F814W  filter
\citep{Koekemoer2007}.  The  median exposure time across  the field is
2028\,s,  which  yields a  limiting  point-source  depth of  27.2\,mag
($5\sigma$).

\begin{table*}
\caption{X-ray AGN and galaxy samples}\label{tab_sample}
\begin{center} 
\begin{tabular}{l c c  c c c}
\hline
field    & 0.5-7\,keV         &  ${\cal R}<24$\,mag &  $z$-spec & $0.1<z<0.6$ & $0.6<z<1.2$ \\
         & selected sample    &  sample           &  sample   & sample &sample   \\ 
CDFS     & 490                & 224               & 176       & 45 (1)  & 87 (10) \\
AEGIS-XD & 859                & 414               & 288       & 55 (7)  & 121 (18)\\ 
C-COSMOS & 1477               & 962               & 726       & 138 (16) & 282  (65)\\
\hline
\end{tabular} 
\begin{list}{}{}
\item  The columns  are: (1)  field name;  (2) total  number  of X-ray
sources detected  in the 0.5-7\,keV  (full) band; (3) total  number of
full-band  selected sources  with ${\cal  R}<24$\,mag, where  $\cal R$
stands for  either the MUSYC $R$-band (CDFS),  the $r^{\prime}$ filter
of  the CFHT  Megaprime camera  (AEGIS-XD) or  the  Subaru Suprime-Cam
instrument  $r^+$ band  (C-COSMOS); (4)  number of  full-band selected
sources  with  ${\cal R}<24$\,mag  and  secure spectroscopic  redshift
measurements; (5)  the same as in  column 4 for  the redshift interval
0.1--0.6. The numbers in the the parentheses correspond to X-ray AGN
with SEDs that are best-fit by  the Seyfert or QSO hybrid templates of
Salvato et  al. (2009, 2011). For  these sources the  optical light is
contaminated by  AGN emission and  is therefore not  representative of
the underlying stellar population.  They are exclude from the analysis
when  studying the  AGN  host galaxy  properties  (e.g. stellar  mass,
optical/near-IR colours);  (6) the same  as column 5 for  the redshift
interval 0.6-1.2.
\end{list}
\end{center}
\end{table*}

\section{Rest-frame properties}\label{sec_rest}
This section describes how the rest-frame colours, X-ray luminosities in
the 2-10\,keV band  and absorbing column densities, $N_H$,  of X-ray AGN
are determined.

The    {\sc   kcorrect}    version   4.2    routines    developed   by
\cite{Blanton_Roweis2007}  are  used to  fit  templates to  the
optical photometry of X-ray sources and estimate rest-frame colours in
the   AB  system.  Rest   frame  magnitudes   are  estimated   in  the
\cite{Bessell90}  $U$  and  $B$  passbands and  the  2MASS-$J$  filter
without any atmospheric corrections or detector response included. The
input photometry  to {\sc kcorrect}  was different for each  field. In
the  case  of  the  CDFS  we used  the  MUSYC  $UBVRIzJHK$  broad-band
photometry \citep{Cardamone2010}.   For AEGIS-XD the  CFHT $ugriz$ and
Palomar  WIRC  (Wide-field  Infrared  Camera)  $JK$  \citep{Bundy2006}
photometry was  employed.  In C-COSMOS  fluxes in the  CFHT $u^\star$,
SUBARU    $Vg^+r^+i^+z^+$   \citep{Capak2007},    UKIRT    WFCAM   $J$
\citep{McCracken2010} and  CFHT WIRCAM $Ks$  \citep{Capak2007} filters
were provided  to {\sc kcorrect}.  When  estimating rest-frame colours
we attempt to minimise  k-corrections, which unavoidably depend on the
adopted set  of model  Spectral Energy Distributions.   The rest-frame
magnitude of a  source in a particular filter,  $X$, is estimated from
the photometry  in the waveband  that has effective wavelength  at the
rest-frame of the source close to that of the filter $X$.  Sources are
split into two broad redshift  bins, 0.1-0.6 and 0.6-1.2. The observed
photometric bands  used to determine  rest frame $UVJ$  magnitudes for
the   sources   in   each   redshift   bin   are   listed   in   Table
\ref{tab_restfilter}.

\begin{table*}
\caption{Observed to rest-frame photometry}\label{tab_restfilter}
\begin{center} 
\begin{tabular}{c   c c  c c  c c}

\hline  
 
rest-frame filter & \multicolumn{6}{c}{Observed band used to estimate
  rest-frame magnitudes for each field  and redshift sub-sample} \\ 
                  &  \multicolumn{2}{c}{CDFS} & \multicolumn{2}{c}{AEGIS-XD} &  \multicolumn{2}{c}{C-COSMOS} \\
                  & $0.1<z<0.6$ & $0.6<z<1.2$ & $0.1<z<0.6$ & $0.6<z<1.2$ & $0.1<z<0.6$ & $0.6<z<1.2$ \\

      Johnson $U$ &  MUSYC $V$  & MUSYC $R$   & CFHT $g$    & CFHT $r$    & Subaru $g^+$& Subaru $r^+$ \\     
      Johnson $V$ &  MUSYC $I$  & MUSYC $z$   & CFHT $i$    & CFHT $z$    & Subaru $i^+$& Subaru $z^+$ \\     
      2MASS   $J$ &  MUSYC $H$  & MUSYC $K$   & WIRC $K$    & WIRC $K$    & WIRCAM $Ks$&  WIRCAM $Ks$ \\     

\hline
 
\end{tabular} 
\begin{list}{}{}
\item 
Listed are the observed bands in  each field that are used to estimate
rest-frame  $UVJ$ magnitudes.  The sources  are split  into  two broad
redshift bins.  At the mean redshift  of each bin  the listed observed
bands have rest-frame effective wavelengths  that are close to those of
the  $U$, $V$  or  $J$  filters. For  the  $0.1<z<0.6$ subsamples  the
observed $H$-band  photometry is best-suited for  the determination of
rest-frame $J$-band  magnitudes.  However, $H$-band  photometry is not
available in the AEGIS-XD and  C-COSMOS fields. We therefore choose to
use the  $K$-band photometry in  those fields to  determine rest-frame
$J$ band magnitudes.  
\end{list}
\end{center}
\end{table*}

The  intrinsic  column density,  $N_H$,  of  individual  X-ray AGN  is
determined from the hardness  ratios between the soft (0.5-2\,keV) and
the hard (2-7\,keV) X-ray  bands assuming an intrinsic power-law X-ray
spectrum  with  index  $\Gamma=1.9$  \citep[e.g.][]{Nandra1994}.   The
derived column densities  are then used to convert  the count-rates in
the 0.5-7\,keV  band to rest-frame 2-10\,keV luminosity,  $L_X(\rm 2 -
10 \, keV)$. We limit the  sample to sources brighter than $L_X \rm (2
- 10\,keV) =  10^{41} \, erg \, s^{-1}$.   Contamination by non-AGN at
faint  luminosities is  a  potential source  of  bias.  Normal  galaxy
candidates are  selected to have $L_X  \rm (2 - 10\,keV)  < 10^{42} \,
erg \,  s^{-1}$, $\rm \log  N_H<22$ ($\rm cm^{-2}$)  and $f_X/f_{{\cal
R}}<-1.5$,  where $f_{\cal  R}$  is  the optical  flux  in the  ${\cal
R}$-band  filter of each  survey field.   Variants of  these selection
criteria  are  often  used  to  identify  normal  galaxies  at  X-rays
\citep[e.g.][]{Georgakakis2007}.  A total of 53 galaxy candidates
are identified among the spectroscopic X-ray selected sample listed in
Table 1.  These sources are removed from the analysis.

Stellar masses  of AGN  host galaxies in  the 3 fields  are calculated
using the methods  presented in \cite{Perez2008} and \cite{Barro2011a,
Barro2011b}. The observed  SED of each source is fit  with a large set
of  templates  based  on   {\sc  pegase}  version  1  \citep{Fioc1997}
tau-models  (running from  a single  stellar population  to continuous
SFR) and assuming a Salpeter  Initial Mass Function (IMF; stellar mass
range  $\rm  0.1  -  100\,M\odot$), different  metallicities  and  the
\cite{Calzetti2000} extinction law.  We  do not measure stellar masses
for  X-ray sources  for which  the  SED fitting  process described  in
Section \ref{sec_xagn} suggests a significant AGN component that could
contaminate the host  galaxy emission. These are sources  fit with any
of  the Seyfert  or QSO  hybrid templates  of Salvato  et  al.  (2009,
2011). The number of these X-ray AGN are listed in Table 1.

\begin{figure*}
\begin{center}
\includegraphics[height=0.75\columnwidth]{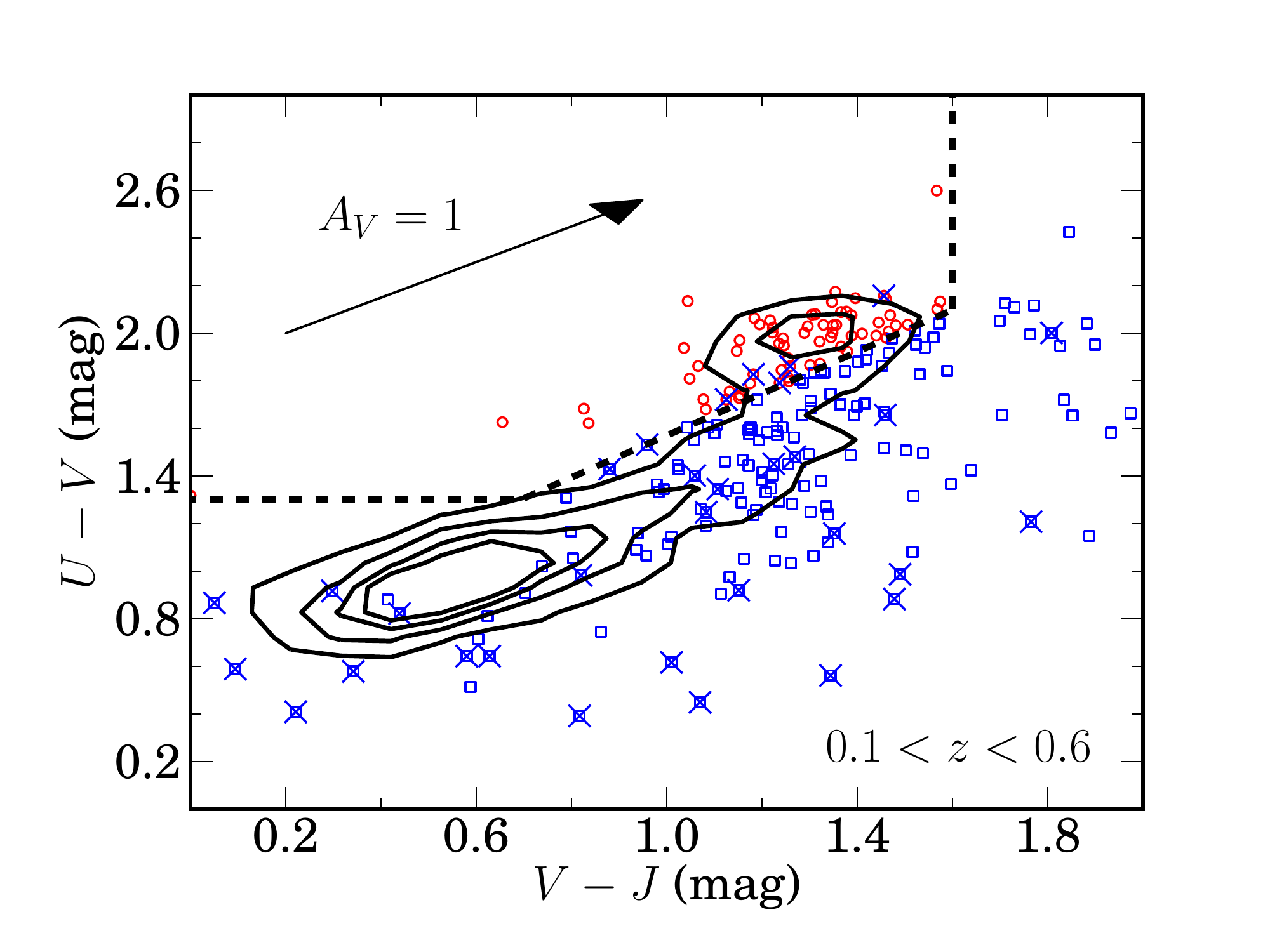}
\includegraphics[height=0.75\columnwidth]{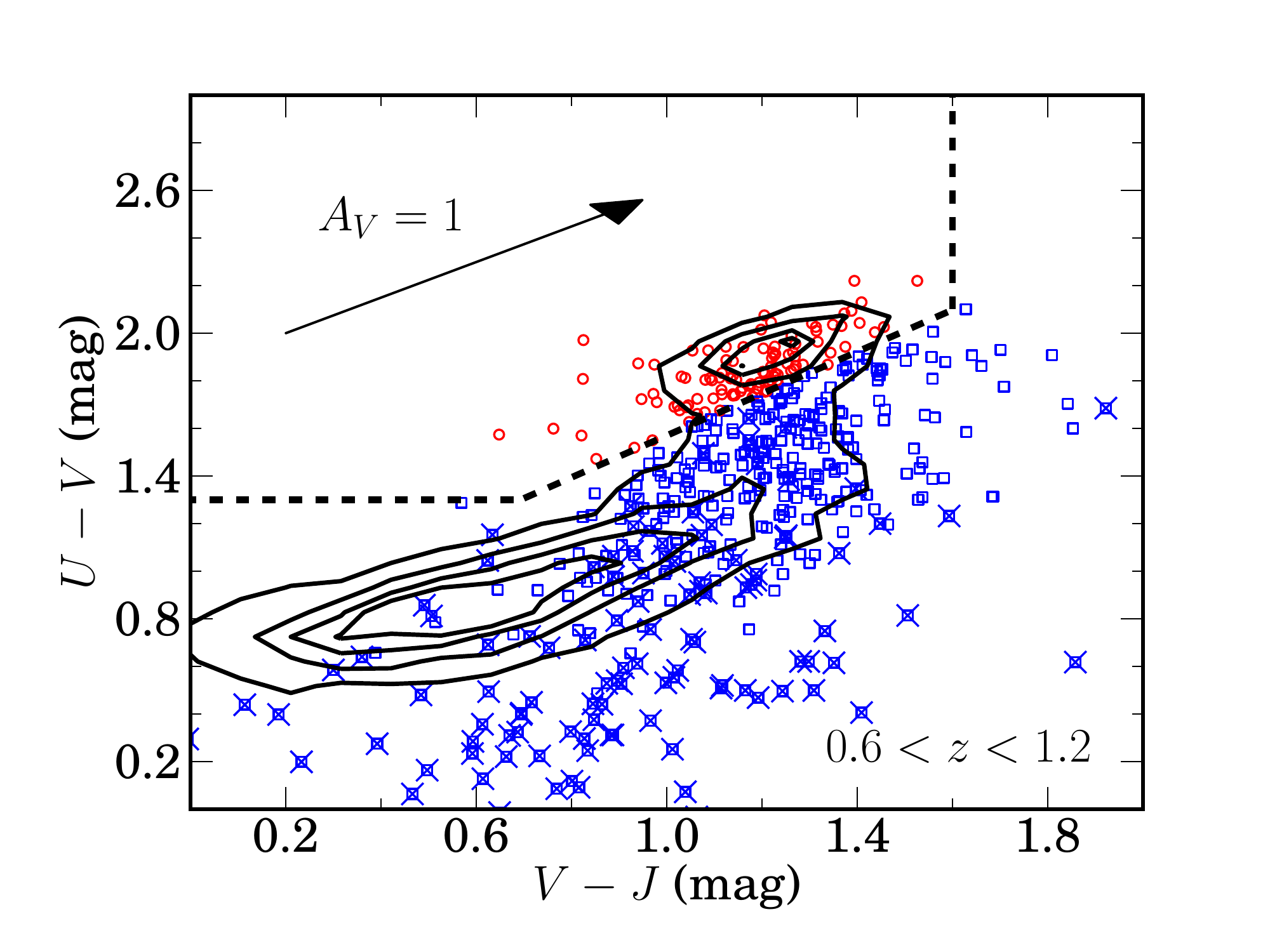}
\end{center}
\caption{$U-V$  vs $V-J$  diagram  of X-ray  AGN  (coloured symbols)  and
galaxies (black contours).  The  galaxy sample in both panels consists
of  sources  with secure  redshifts  obtained  as  part of  the  large
spectroscopic follow-up  campaigns in the CDFS,  AEGIS-XD and C-COSMOS
fields, e.g.  DEEP2, DEEP3  and the VIMOS/zCOSMOS bright project.  The
different contour  levels correspond  to 30, 60,  90 and  120 galaxies
within  bins  of  size  0.1\,mag.   The  dashed  lines  correspond  to
$U-V=0.88\,(V-J)+0.69$, $U-V>1.3$, $V-J<1.6$  (Williams et al.  2009).
In both panels galaxies are distributed into two distinct populations,
i.e. quiescent  and star-forming.  The wedge, as  defined above, marks
the transition region between  these two galaxy populations. The arrow
shows  the reddening  vector  with  $A_V=1$ for  the  Calzetti et  al.
(2000) law.  This is parallel to the quiescent galaxy selection wedge.
Dusty  star-forming galaxies  are therefore  separated  from quiescent
systems. Red circles  are X-ray AGN in the quiescent  region of the $UVJ$
diagram.   Blue squares   are  X-ray  in the  star-forming  part  of  the
colour-colour space.  Crosses on top  of an X-ray AGN mark sources for
which AGN radiation contaminates the underlying host galaxy continuum.
The   rest-frame  colours   of   those  sources   are  therefore   not
representative of their hosts.  }\label{fig_UVJ}
\end{figure*}

\section{The AGN X-ray luminosity function}\label{sec_xlf}

The X-ray  luminosity function  of AGN is  derived using  the standard
non-parametric  $\rm 1/V_{max}$  method \citep{Schmidt1968}.   In this
calculation  we take into  account the  X-ray selection  function, the
optical  magnitude limit  of different  samples and  the spectroscopic
identification  incompleteness.   The   XLF  in  logarithmic  bins  is
estimated by the relation

\begin{equation} \phi(L_X, z) \, dL_X = \sum_{i} \frac{w_{i}}{V_{max,i}},
\end{equation}

\noindent   where   $w_{i}$   is    the   weight   applied   to   each
spectroscopically   identified   source  $i$   to   correct  for   the
spectroscopic incompleteness  (see below). $V_{max,i}$  is the maximum
comoving  volume  for  which  the  source  $i$  satisfies  the  sample
selection criteria,  i.e.  redshift range,  apparent optical magnitude
limit and X-ray flux  limit.  $V_{max,i}$ depends on X-ray luminosity,
absolute  optical magnitude,  redshift as well  as  the overall
shape of the optical and X-ray SED

\begin{equation}\label{eq_vmax_x}
 V_{max,i}   =   \frac{c}{H_0}   \int_{z1}^{z2}   \,   \Omega(L_X,N_H,z)\,
 \frac{dV}{dz}\,dz\, dL,
\end{equation}

\noindent where  $dV/dz$ is the  volume element per  redshift interval
$dz$.   The integration limits  are $z1=z_L$  and $z2=min(z_{optical},
z_U)$,  where we  define $z_L$,  $z_U$  the lower  and upper  redshift
limits applied  to the sample and $z_{optical}$  is the redshift
at which the source becomes  fainter than the survey optical magnitude
limit.  $\Omega(L_X,N_H,z)$ is the solid angle of the X-ray survey
available to a  source with luminosity $L_X$ and  column density $N_H$
at  a  redshift  $z$ (corresponding  to  a  flux  $f_X$ on  the  X-ray
sensitivity curve). The uncertainty at  a given luminosity or mass bin
is

\begin{equation} 
\delta \phi^2 = \sum_{i} \left ( \frac{w_{i}}{V_{max,i}} \right )^2.
\end{equation}

\noindent The conversion of the absolute to apparent optical magnitude
in the 1/Vmax calculation uses the optical k-corrections determined by
the  {\sc kcorrect}  version 4.2  routines \citep{Blanton_Roweis2007}.
The model that  best fits the optical photometric data  of a source is
also used to  estimate k-corrections for the same  source at different
redshifts. In the  case of the XLF, the  intrinsic $N_H$ of individual
X-ray  sources is  taken into  account in  the 1/Vmax  estimation. The
X-ray k-corrections  are calculated by adopting  an absorbed power-law
spectral  energy  distribution  with  $\Gamma=1.9$  and  photoelectric
absorption  cross  sections as  described  by \cite{Morrison1983}  for
solar metallicity.

The weight $w_i$ is estimated  following a methodology similar to that
described  by \cite{Lin1999} and  \cite{Willmer2006}.  For  each X-ray
source, $i$, in  the sample we estimate the  probability $P_i$ that it
lies   within   the   redshift   interval  of   interest   $z_L<z<z_U$
(e.g. $0.6<z<1.2$). Spectroscopically  identified sources are assigned
$P_i=1$  if $z_L<z<z_U$ or  else $P_i=0$.   For X-ray  sources without
spectroscopic redshifts  we integrate the photometric  redshift PDZ to
determine $P_i$.

We  then define a  three dimensional  observed colour-colour-magnitude
space.   For each  source with  secure spectroscopic  redshift  in the
range $z_L<z<z_U$, we sum the  probabilities $P_i$ of all nearby X-ray
sources  within  a  colour-colour-magnitude  sphere. Within  the  same
sphere we  also count the  number of X-ray sources  with spectroscopic
redshifts in  the interval  $z_L<z<z_U$, $N_{spec}$. The  weight $w_i$
for each spectroscopic source  is $\sum_i P_i/ N_{spec}$. Typical
weight values are 1.12 for the redshift range $0.1<z<0.6$ and 1.25 for
the X-ray AGN in the interval $0.6<z<1.2$.

The XLF is estimated  separately in the redshift intervals ($z_L=0.1$,
$z_U=0.6$) and ($z_L=0.6$, $z_U=1.2$). The data spheres are defined by
the  observed $\cal  R-I$, $\cal  I-K$ colours  and the  $\cal R$-band
magnitude.   As in  the previous  section the  symbols $\cal  RIK$ are
defined as MUSYC $RIK$ for  CDFS, CFHT $ri$ and WIRC $K$, respectively
for  AEGIS-XD, SUBARU $r^+i^+$  and WIRCAM  $Ks$, respectively  in the
case of C-COSMOS.

\begin{figure*}
\begin{center}
\includegraphics[height=0.9\columnwidth]{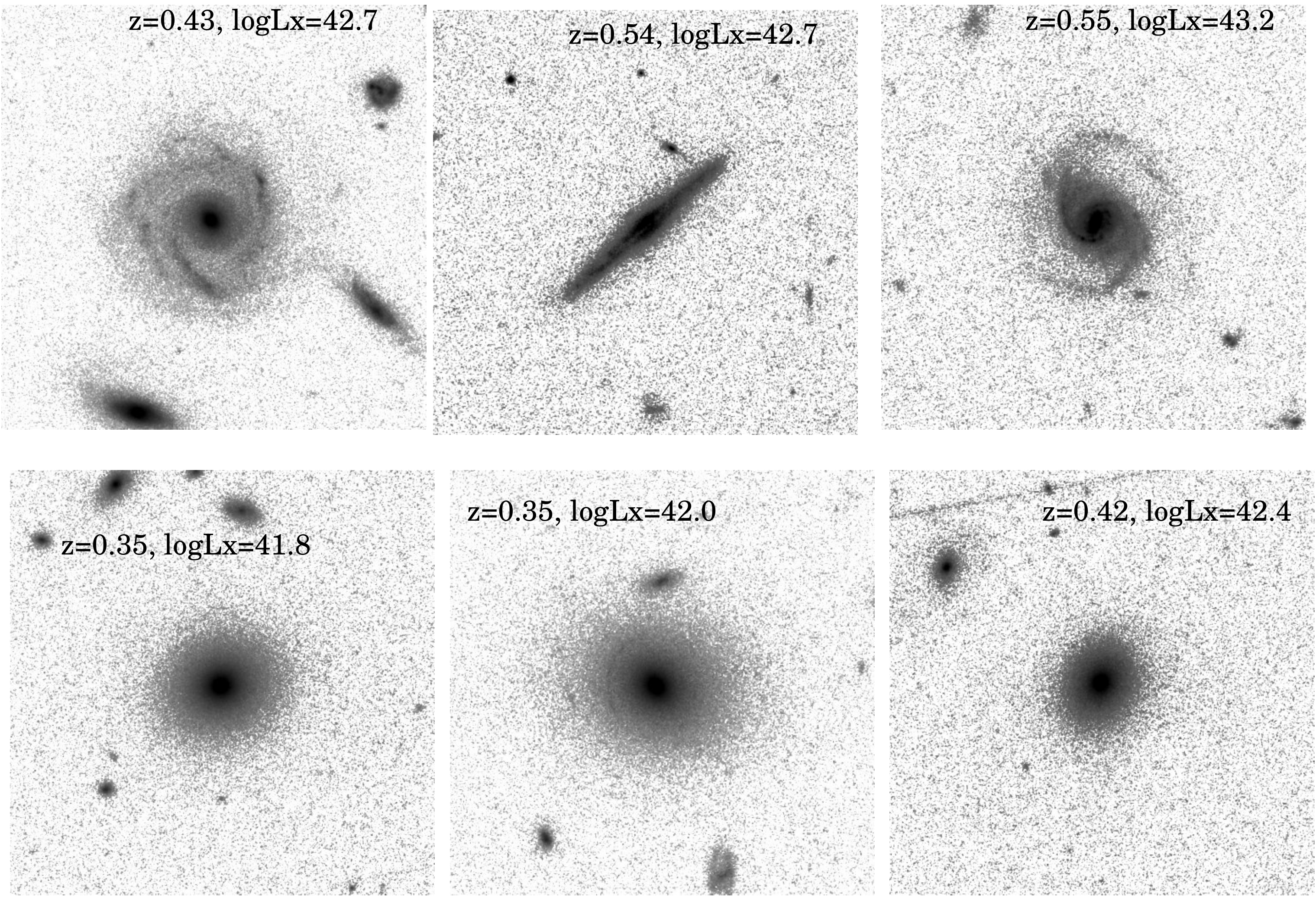}
\end{center}
\caption{Examples of HST/ACS morphologies of X-ray AGN in the C-COSMOS
  field  associated with galaxies  in the  quiescent (bottom  row) and
  star-forming (top row) region of  the $UVJ$ diagram.  The images are
  15\,arcsec  on the  side  and  have a  pixel  scale of  0.03\,arcsec
  \citep{Koekemoer2007}.  The filter used in the HST/ACS survey of the
  COSMOS field is the F814W.  }\label{fig_morph_qual}
\end{figure*}

\begin{figure}
\begin{center}
\includegraphics[height=0.9\columnwidth]{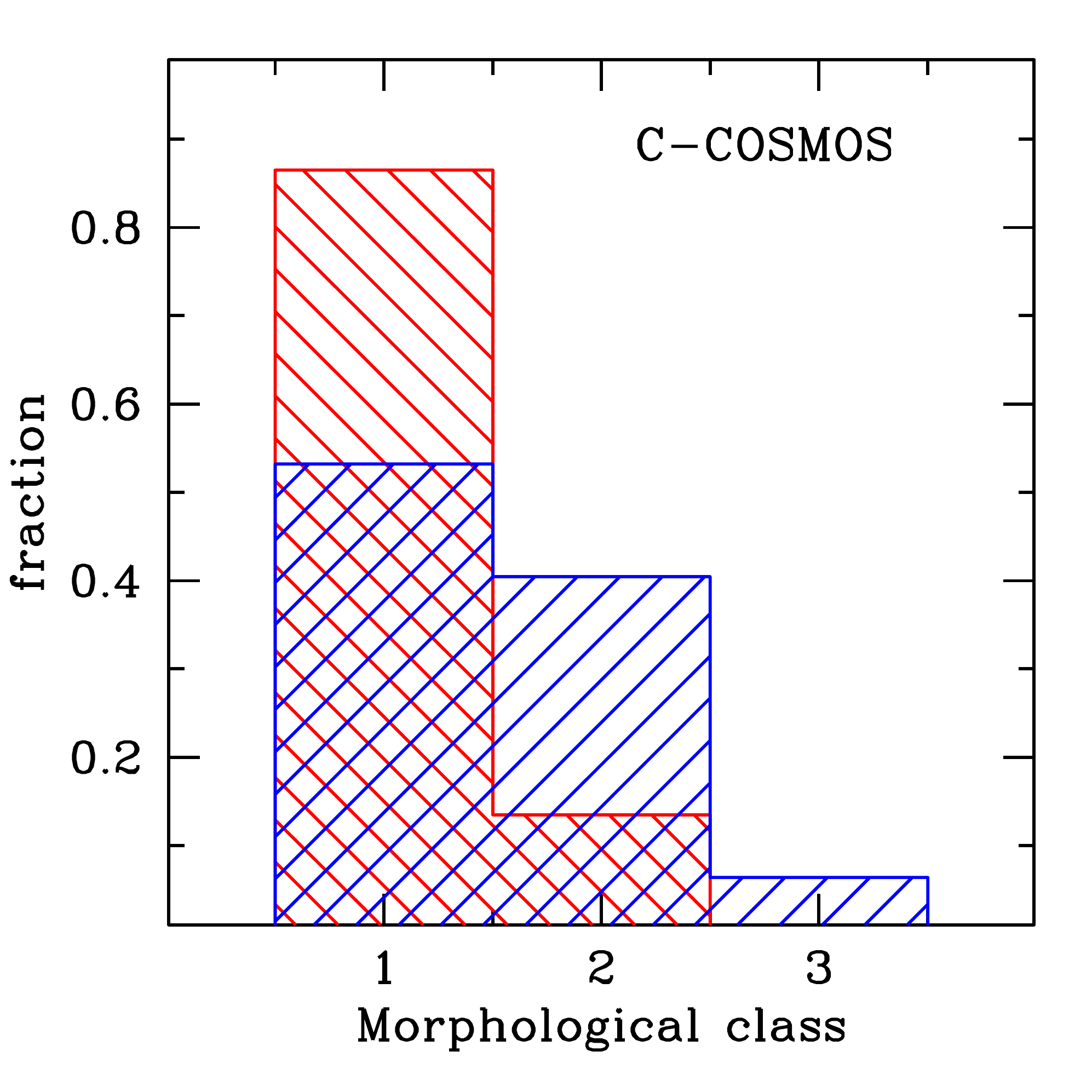}
\end{center}
\caption{Morphological mix  of X-ray AGN  hosts in the  C-COSMOS field
(spectroscopic redshift  interval 0.6-1.2).  The sample is  split into
three groups, early-types, spirals and irregulars, which correspond to
morphological  class  numbers  1,  2  and  3  respectively  (Tasca  et
al. 2009).  The red and  blue histograms correspond to X-ray AGN hosts
in  the  quiescent  and  star-forming  region of  the  $UVJ$  diagram.
}\label{fig_morph_cosmos}
\end{figure}

\begin{figure}
\begin{center}
\includegraphics[height=0.9\columnwidth]{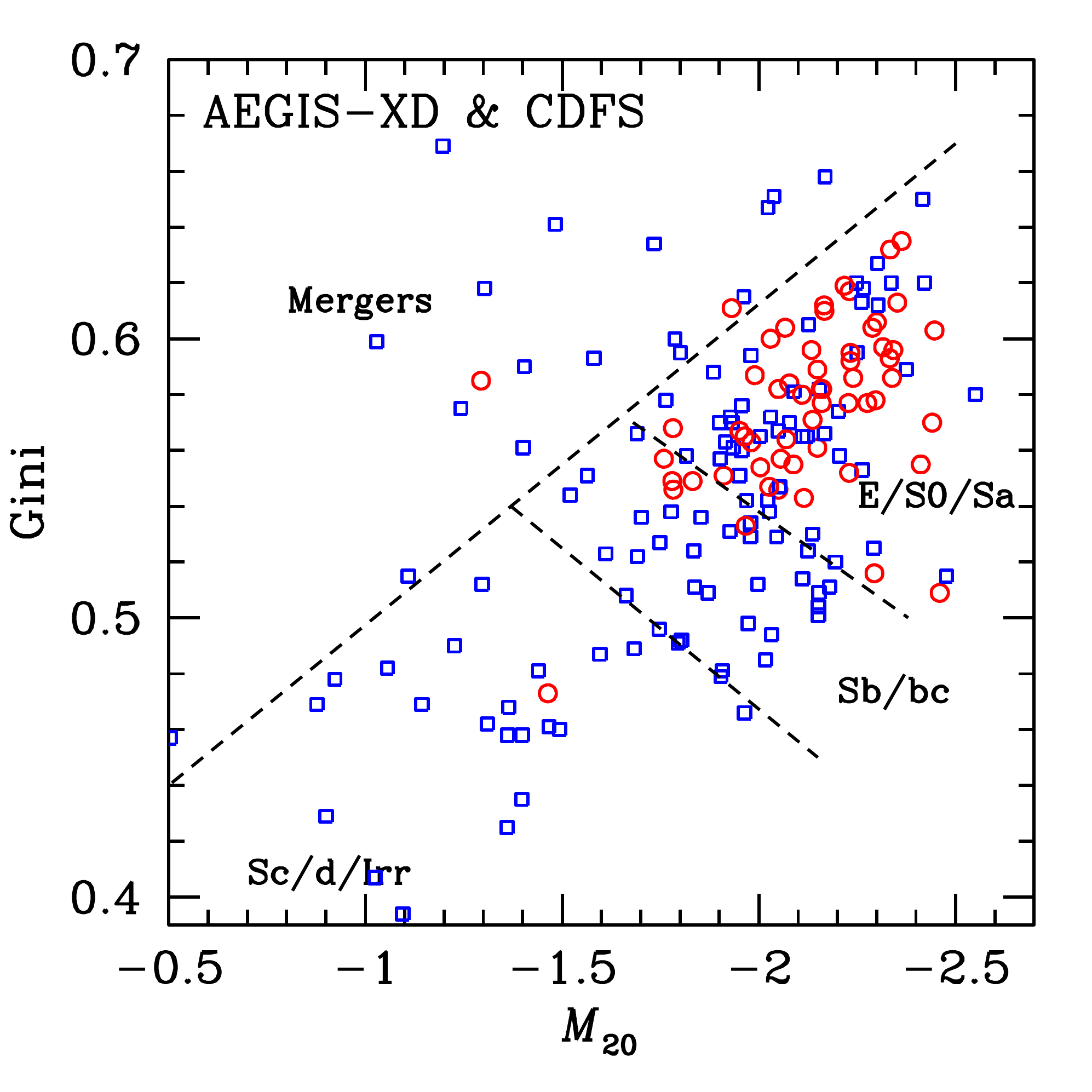}
\end{center}
\caption{Gini-$M_{20}$ diagram for X-ray  AGN in the AEGIS-XD and CDFS
fields  (spectroscopic redshift  interval 0.6-1.2) with counterparts
  in the HST/ACS surveys of those fields 
\citep{Lotz2008_aegis, Messias2011phd}.  The  regions of the parameter
space  occupied by  different  galaxy types  are  demarcated with  the
dashed lines.   Red circles and  blue squares correspond to  X-ray AGN
hosts in  the quiescent and  star-forming region of the  $UVJ$ diagram
respectively.  The Gini and $M_{20}$ parameters are estimated from the
HST/ACS    images    in   F814W    (AEGIS-XD)    and   F775W    (CDFS)
filters}\label{fig_morph_aegis}
\end{figure}

\begin{figure*}
\begin{center}
\includegraphics[height=0.9\columnwidth]{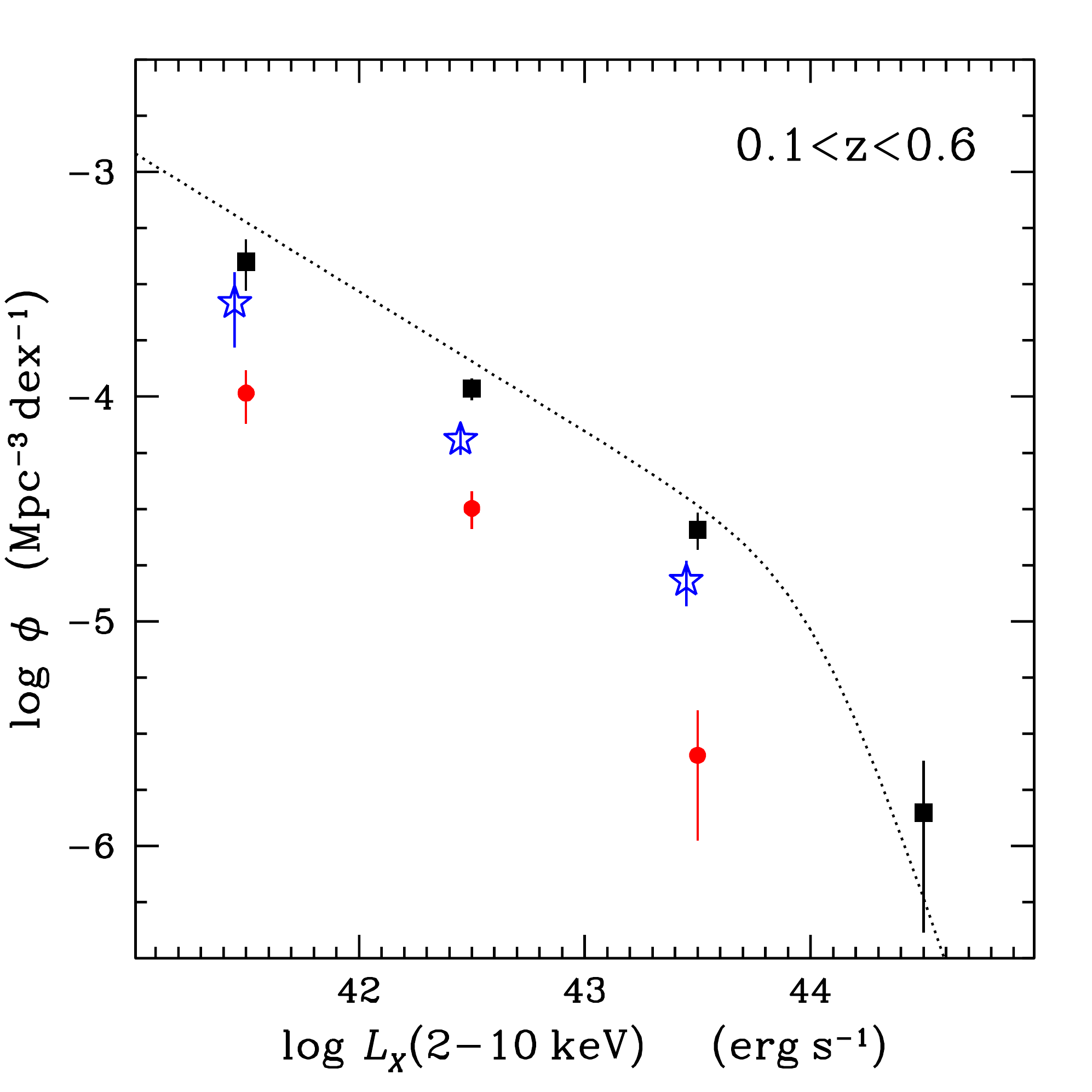}
\includegraphics[height=0.9\columnwidth]{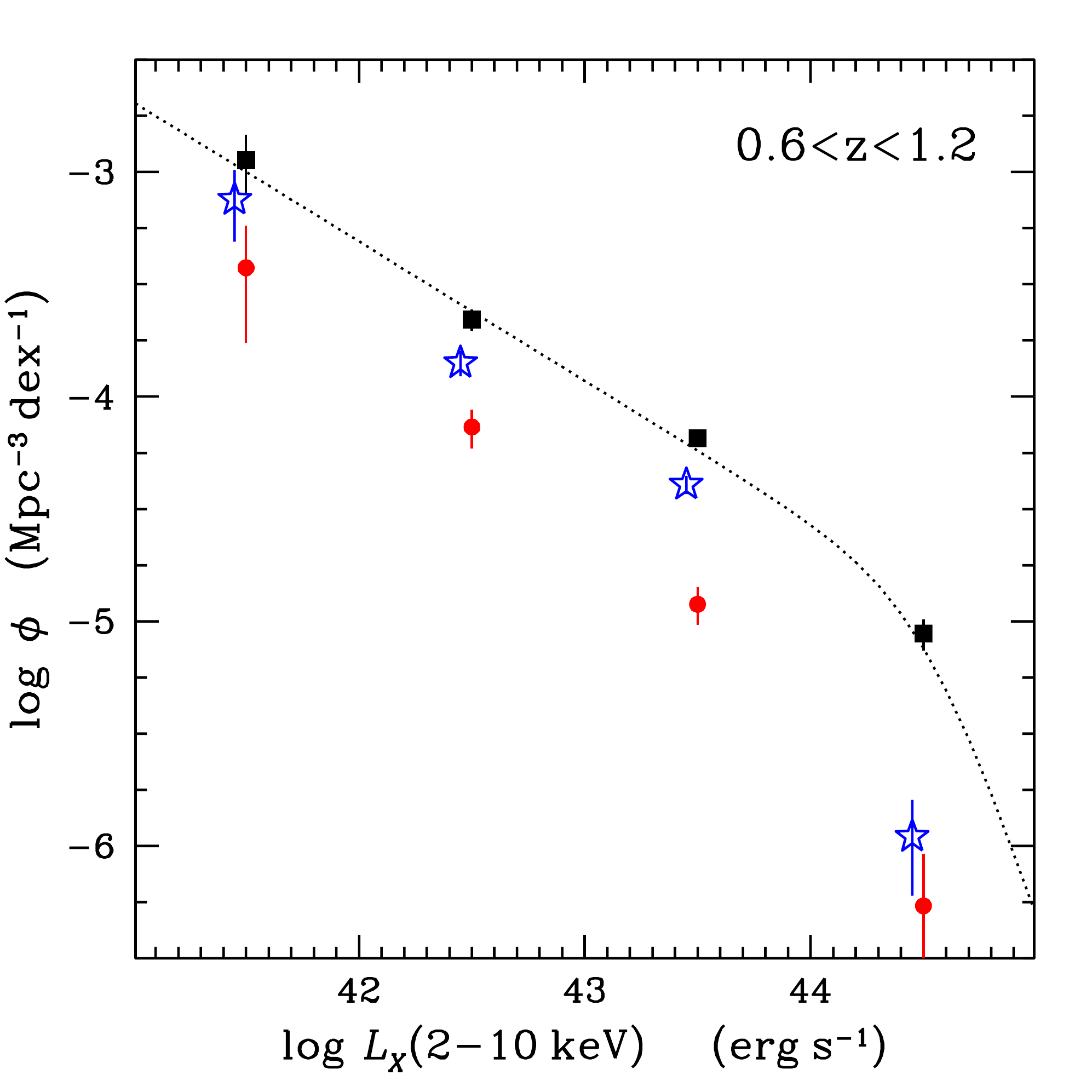}
\end{center}
\caption{The  2-10\,keV X-ray luminosity  function. The  black squares
  are the XLF estimates from  the combined CDFS, AEGIS-XD and C-COSMOS
  fields in  the redshift intervals $z=0.1-0.6$  and $z=0.6-1.2$.  The
  dotted  lines correspond  to  the Luminosity  And Density  Evolution
  (LADE) model of Aird et  al. (2010) estimated at the median redshift
  of each  sample. In  all panels  the red circles  are AGN  hosted by
  galaxies in the quiescent wedge of the $UVJ$ diagram. Blue stars are
  for AGN associated with  star-forming galaxies (both blue and dusty)
  in the  $UVJ$ diagram.  Systems  in which AGN  emission contaminates
  the host galaxy colours are not plotted.}\label{fig_xlf}
\end{figure*}

\section{Results}

\subsection{Star-formation properties of X-ray AGN hosts}

Figure \ref{fig_UVJ}  presents the $UVJ$  diagram of X-ray AGN  in two
redshift bins, 0.1--0.6 and 0.6--1.2.   For comparison we also plot in
the  same figure  the  rest-frame $UVJ$  colours of  spectroscopically
confirmed galaxies  in the MUSYC,  AEGIS-XD and C-COSMOS  fields using
the   methodology   described    in   section   \ref{sec_rest}.    The
spectroscopic  redshifts of  the  galaxy samples  are  from the  MUSYC
compilation, DEEP2,  DEEP3 \citep{Newman2012, Cooper2011_deep3_1,
  Cooper2012_deep3_2} and  the VIMOS/zCOSMOS bright  project
\citep{Lilly2009}.  

In   Figure  \ref{fig_UVJ}  quiescent   systems  are   separated  from
star-forming (including dusty) galaxies by the selection wedge defined
by  the  relations  $U-V>1.3$,  $V-J<1.6$  and  $U-V>0.88\,(V-J)+0.69$
\citep{Williams2009}.  The specific star-formation rate of galaxies is
found  to change  rapidly across  the  wedge, at  least for  redshifts
$z\la1.5$ \citep{Williams2009}.  Some level  of mixing between low and
high  specific  star-formation  rate   galaxies  is  expected  at  the
transition region  of the $UVJ$  diagram.  Nevertheless, to  the first
approximation   the  $UVJ$  colour-colour   plot  provides   a  simple
diagnostic of  the level of star-formation in  galaxies.  Turning next
to X-ray AGN, they are found in galaxies in both the quiescent and the
star-forming region  of Figure \ref{fig_UVJ}.  This  suggests that the
growth of SMBHs to $z\approx1$ is taking place in galaxies with a wide
range of  star-formation histories. The  apparent displacement in
Figure \ref{fig_UVJ} between star-forming  galaxies (peak of the black
contours)  and X-ray AGN  hosts in  the star-forming  part of  the UVJ
diagram  (blue  squares) is  because  of  the  different stellar  mass
distributions  of  the  two  populations.   X-ray  AGN  are  typically
associated   massive  hosts,  while   star-forming  galaxies   in  any
magnitude-limited sample inlcude a  large fraction of low stellar mass
systems.

The morphology of  AGN hosts also changes across  the quiescent galaxy
selection wedge of Figure \ref{fig_UVJ}.  This underlines that the
distribution of X-ray AGN on the $UVJ$ diagram reflects differences in
the properties of their hosts. X-ray AGN in the quiescent wedge of the
$UVJ$ diagram  are dominated by  ellipticals.  In contrast,  X-ray AGN
hosts in  the star-forming part of  the $UVJ$ diagram  include a large
fraction of spirals.  Examples of the morphologies of AGN hosts in the
C-COSMOS field are  presented in Figure \ref{fig_morph_qual}.  Figures
\ref{fig_morph_cosmos}  and  \ref{fig_morph_aegis}  plot  quantitative
non-parametric measures of the host galaxy morphology for X-ray AGN in
the C-COSMOS, AEGIS-XD and CDFS  fields. In the C-COSMOS field we use
the  morphological  catalogue  of  \cite{Tasca2009}.  They  classified
galaxies detected in the HST/ACS survey of the COSMOS field into early
types,  spirals  and  irregulars   based  on  their  position  in  the
multi-dimensional space  defined by the  galaxies' apparent magnitudes
and three  non-parametric morphological quantities,  the Concentration
index,   the   asymmetry    parameter   and   the   Gini   coefficient
\citep{Abraham2003, Lotz2004}. Figure \ref{fig_morph_cosmos} shows the
morphological mix of X-ray AGN  hosts in the C-COSMOS survey.  Sources
in  the quiescent  region of  the $UVJ$  diagram are  mostly  found in
bulge-dominated  hosts ($\approx$85\%)  and only  a small  fraction is
associated with  disks ($\approx$15\%).  In contrast X-ray  AGN in the
blue  part of  the  $UVJ$  diagram are  nearly  equally split  between
early-types and disks/irregulars.  Similar results are obtained in the
AEGIS-XD and  CDFS fields.  The morphologies of  the galaxies detected
in  the HST/ACS surveys  of those  fields are  quantified by  the Gini
coefficient and the second moment  of the brightest 20\% pixels of the
galaxy,  $M_{20}$  \citep{Lotz2008_aegis, Messias2011phd}.   Different
Hubble  types are  separated  in the  Gini--$M_{20}$  diagram and  the
morphological  classification   based  on  these   two  non-parametric
estimators    remains    robust     to    high    redshift.     Figure
\ref{fig_morph_aegis}  shows that  X-ray  AGN hosts  in the  quiescent
region of the  $UVJ$ diagram are distributed in  the early-type region
of the  Gini--$M_{20}$ parameter space.  In contrast  a large fraction
of X-ray AGN with blue $UVJ$ colours scatter into the late-type region
of the Gini--$M_{20}$ diagram.

The evidence above is consistent with different physical conditions of
black hole growth in AGN selected on the basis of the $UVJ$ colours of
their hosts.  This perhaps reflects diverse accretion modes, which can
be  isolated   by  selecting  on   the  star-formation  rate   of  AGN
hosts. Splitting AGN  samples by the level of  star-formation of their
hosts could  therefore place limits  on the significance  of different
fuelling  modes to  the  accretion history  of  the Universe.   Figure
\ref{fig_xlf}  for example,  shows the  2-10\,keV  XLF of  AGN in  the
redshift intervals  $z=0.1-0.6$ and 0.6--1.2 split  into quiescent and
star-forming hosts based on their  position in the $UVJ$ diagram.  The
latter population dominates the space  density of AGN at both redshift
intervals.  Nevertheless,  active black holes in  quiescent hosts also
have a non-negligible  contribution to the XLF. This  result is placed
into a  quantitative footing by estimating for  each redshift interval
the  integrated  X-ray  luminosity  density  associated  with  AGN  in
quiescent  and star-forming hosts  based on  their $UVJ$  colours.  We
then normalise to the total  X-ray luminosity density in each redshift
bin and plot the results against look-back time and redshift in Figure
\ref{fig_ld_fraction}. The accretion  density is dominated by actively
star-forming galaxies at redshifts 0.1--1.2.  AGN in quiescent systems
have a small, but non-negligible, contribution to the X-ray luminosity
density, $\approx15-20$ per cent.   It is also interesting that within
the  errors,  the  accretion  density in  quiescent  and  star-forming
galaxies does  not appear  to evolve strongly  in the last  8\,Gyrs of
cosmic time.

We also  caution that  a fraction of  the accretion density  in Figure
\ref{fig_ld_fraction} is  associated with X-ray sources  for which the
AGN  light  dominates  the  observed UV/optical  continuum.  For  this
population we have  no handle on the level  of star-formation of their
hosts  because  their $UVJ$  colours  are  not  representative of  the
underlying  stellar  population.  Studies  of  broad-line  AGN at  low
redshift  ($z\la   0.1$)  suggest  that  they  are   mostly  found  in
star-forming hosts \citep{Trump2013}.  In  our work however, we prefer
to keep these sources as a separate class. They are identified via the
template SED fits to  the observed multi-waveband photometry described
in  section  \ref{sec_xagn}.   X-ray  AGN  that are  best-fit  by  the
AGN/galaxy  hybrid templates  of Salvato  (2009, 2011)  are  marked as
potentially  having $UVJ$  colours  contaminated by  the central  AGN.
Figure \ref{fig_UVJ} shows that  this approach identifies the majority
of X-ray sources with very blue  $U-V$ colours.  The
preference  for  an AGN/galaxy  hybrid  template  by  the SED  fitting
process correlates well  with the presence of broad  emission lines in
the optical spectra of individual sources \citep[][]{Lusso2012}.

\subsection{Specific accretion rate}

If  the  level  of   star-formation  of  AGN  hosts  traces  different
conditions of  black hole growth,  one may also expect  differences in
the accretion properties  of the SMBH as a  function of star-formation
rate.   It is  therefore interesting  to explore  the  Eddington ratio
distribution ($\lambda_{Edd}$,  observed accretion rate  onto the SMBH
relative  to  the  Eddington  limit)  between  AGN  in  quiescent  and
star-forming hosts. The Eddington ratio relates directly to properties
of the active black hole and  is therefore the quantity one would like
to  study  in relevance  to  host  galaxy  properties.  This  exercise
however, is  limited by the ability  to measure the mass  of the black
hole of individual AGN in the absence of broad optical emission lines,
e.g.  because of obscuration.  In this case, one has to estimate first
the bulge  mass of  the host galaxy  and then assume  a Magorrian-type
scaling relation  to approximate the  mass of the central  black hole.
Both steps however,  are not trivial and may  suffer uncertainties and
systematics, particularly in the case of high redshift AGN samples.

The specific  accretion rate, $\lambda$, defined as  the ratio between
the   AGN  accretion   luminosity   and  host   galaxy  stellar   mass
\citep{Aird2012, Bongiorno2012}, is advantageous because it is related
to quantities that can  be measured with systematic uncertainties that
are  typically   smaller  than  in   the  case  of  black   hole  mass
determinations.  The specific accretion rate measures how fast a black
hole grows  relative to the  integrated star-formation history  of its
host.  For bulge dominated galaxies,  $\lambda$ is also a proxy of the
Eddington  ratio.   The  next  section discusses  differences  between
$\lambda$ and $\lambda_{Edd}$.

When  constructing the  $\lambda$ distribution  of AGN  there  are two
selection biases that need to  be accounted for.  The first relates to
the  fact that  the  AGN  sample is  optical  magnitude limited,  i.e.
${\cal R}<24$\,mag.  This translates  to different stellar mass limits
for  star-forming  and  quiescent   hosts  because  of  the  different
mass--to--light  ratio  of   their  stellar  populations.   This
introduces incompleteness as passive  galaxies of a given stellar mass
drop  out of  the sample  at lower  redshift compared  to star-forming
galaxies   of  the   same  stellar   mass.    Figure  \ref{fig_massz}
demonstrates  this  source of  bias  by  plotting  stellar mass  as  a
function of redshift  for AGN hosts colour-coded by  their position on
the $UVJ$ diagram.  Quiescent hosts  are scarce at low stellar masses.
We minimise this source of  bias by applying a redshift-dependent mass
limit   which  corresponds   to   a  maximally   old  (i.e.    maximal
mass--to--light  ratio)  galaxy.   This  is  defined  by  a  passively
evolving stellar  population that formed by an  instantaneous burst at
$z=5$.  We use the \cite{BC03}  model with a Salpeter IMF to construct
the evolving  SED of  such a stellar  population and estimate  at each
redshift the stellar mass that corresponds to an observed magnitude of
${\cal R}=24$\,mag (see Fig.  \ref{fig_massz}).  Above this mass limit
the  galaxy sample  is not  affected  by incompleteness  as no  galaxy
should have a mass--to--light ratio greater than that of the maximally
old stellar population model.

Another source of  bias is related to the  minimum X-ray luminosity we
adopt for  identifying AGN  among galaxies.  AGN  in low  stellar mass
hosts are  detected above the $L_X$  limit of the sample  only if they
have higher specific accretion rates compared to AGN in higher stellar
mass galaxies.   We account for this  bias by applying  a minimum host
galaxy  stellar mass  limit.  This  translates to  a  minimum specific
accretion rate  below which incompleteness  is kicking in.   We choose
minimum stellar masses of  $\rm M_{star}> 10^{10}\,M_{\odot}$ and $\rm
>10^{11}  \,  M_{\odot}$  for  AGN  at redshifts  $<0.6$  and  $>0.6$,
respectively.   Figure  \ref{fig_massz}  shows  that this  choice,  in
combination with the mass limit of a maximally old stellar population,
result in nearly volume limited  AGN samples in the redshift intervals
0.1--0.6 and 0.6--1.0. These two  subsamples are used to construct and
compare the  specific accretion  rate distributions of  AGN associated
with host  galaxies in the  quiescent and star-forming regions  of the
$UVJ$  diagram.    The  total   number  of  $UVJ$   passive  and
  star-forming AGN hosts are respectively 43, 81 ($0.1<z<0.6$) and 50,
  91 ($0.6<z<1$).

The  specific  accretion  rate  is  estimated  as  the  ratio  of  the
bolometric accretion luminosity, $L_{bol}$, of the AGN and the stellar
mass  of  its host,  $M_{\star}$  (see  section \ref{sec_rest}).   The
$L_{bol}$ is estimated from the X-ray luminosity in the 2-10\,keV band
by  adopting the bolometric  corrections of  \cite{Marconi2004}.  For
the  construction of  the specific  accretion rate  distributions each
source $i$ is weighted by  the same factor used to estimate luminosity
function, i.e.   $w_i/V_{max,i}$ (see section  \ref{sec_xlf}). Sources
in  the   sample  for  which   the  underlying  stellar   emission  is
contaminated  by AGN  light  are  not used  in  the analysis.   Figure
\ref{fig_edd}  plots the space  density of  AGN in  specific accretion
rate  bins for the  redshift intervals  $0.1-0.6$ and  $0.6-1.0$.  The
upper  x-axis  in  both  panels  of  Figure  \ref{fig_edd}  shows  the
conversion between  specific accretion rate and  Eddington ratio under
the     assumptions    of     a    bulge-dominated     galaxy    (i.e.
$M_{star}=M_{bulge}$)  and a  bulge  mass to  black-hole mass  scaling
relation of $M_{SMBH}=0.002\,M_{bulge}$ \citep{Marconi_Hunt2003}.

The  statistical  methodology based  on  the Kolmogorov--Smirnov  test
presented in  the Appendix is  used to compare the  specific accretion
rate distributions of AGN  in star-forming and quiescent hosts plotted
in Figure  \ref{fig_edd}.  We  estimate a null  hypothesis probability
that the  two samples are drawn  from the same parent  population of 5
and  25 per cent  for AGN  in the  redshift intervals  $0.1<z<0.6$ and
$0.6<z<1.0$,  respectively.   The comparison  is  limited to  specific
accretion rates  above the completeness limits,  i.e.  vertical dotted
lines, of Fig. \ref{fig_edd}.  We therefore find evidence, significant
at    the    $2\sigma$   level,    that    low    redshift   AGN    in
star-forming/quiescent  hosts have  different specific  accretion rate
distributions.  For the high redshift sub-sample however, the specific
accretion rates  of AGN split by  their position on  the $UVJ$ diagram
are consistent.

\subsection{Eddington ratio vs specific accretion rate}

The specific accretion  rate is a proxy of the  Eddington ratio of AGN
only for  bulge dominated galaxies  under the assumption of  a scaling
relation    between     bulge    mass    and     black    hole    mass
\citep[e.g.][]{Magorrian1998}.    Figures  \ref{fig_morph_cosmos}  and
\ref{fig_morph_aegis}   however,   show  that   about   half  of   the
star-forming X-ray AGN hosts are late-type galaxies in which the bulge
mass  represents a  fraction  of  the total  stellar  mass.  For  this
subsample  the  specific  accretion  rate  likely  underestimates  the
Eddington ratio.

If we were  to estimate the Eddington ratios of  the current X-ray AGN
sample we should substitute the total stellar mass with the bulge mass
of  the  host  galaxy  and  adopt a  Magorrian-type  scaling  relation
\citep[e.g.][]{Marconi_Hunt2003}  to determine the  mass of  the black
hole.  For early-type hosts the total  stellar mass is a good proxy of
the bulge mass and therefore  to the first approximation the Eddington
ratio  differs from  the specific  accretion rate  only by  a constant
(i.e.  upper  x-axis of Fig.  \ref{fig_edd}).   For late-type galaxies
however, there is an additional factor $M_{bulge}/M_{total}$, i.e. the
ratio of the bulge to total stellar mass, that should also be included
in the calculation.   This factor would shift the  Eddington ratios of
X-ray AGN  in late-type  galaxies to higher  values compared  to those
plotted in the upper  x-axis of Figure \ref{fig_edd}.  This correction
could  potentially alter the  overall $\lambda_{Edd}$  distribution of
AGN in star-forming hosts relative to those in quiescent galaxies.

We  explore  this possibility  by  assuming  for  late-type AGN  hosts
$M_{bulge}/M_{total}=0.5$,  i.e.   typical  for  Sb/Sbc-type  galaxies
\citep[e.g.][]{Fukugita1998,    Oohama2009}.    Assuming    a   single
$M_{bulge}/M_{total}$ ratio is clearly an approximation. Late-type AGN
hosts likely  span a  range of  Hubble types and  even within  a given
morphological  class the  bulge  to total  stellar  mass ratio  varies
considerably.     Nevertheless,   the    simplistic    assumption   of
$M_{bulge}/M_{total}=0.5$  for  all  late-type hosts  illustrates  the
direction  and  amplitude  of   the  change  in  the  Eddington  ratio
distribution of  AGN one should  expect once more  accurate black-hole
mass estimates \citep[i.e. factor of few,][]{Shen2013} for individual
AGN become available.

The  black  solid  line  in  Figure \ref{fig_edd}  shows  the  updated
$\lambda_{Edd}$  distribution of  AGN in  star-forming  hosts assuming
$M_{bulge}/M_{total}=0.5$   for  the   late-type   sub-population  and
$M_{bulge}/M_{total}=1$  for  early-types.   For the  construction  of
those distributions we account for the fact that the AEGIS-XD and CDFS
fields have  only partial HST/ACS  coverage. For those fields  we only
consider the  subregion with HST/ACS  data.  The total  number of
star-forming  AGN hosts with  HST/ACS data  in the  redshift intervals
$0.1<z<0.6$  and  $0.6<z<1.0$  are   80  and  87,  respectively.   As
expected, the  overall impact of  using variable $M_{bulge}/M_{total}$
to  approximate  the  black  hole  mass  of  galaxies  with  different
morphologies  is  a  shift  to  higher  Eddington  ratios  of  AGN  in
star-forming  hosts.  The  net effect  is an  increase of  their space
density  at $\log  \lambda_{Edd}\ga-2$  relative to  AGN in  quiescent
galaxies.

We assess differences in  the Eddington ratio distributions plotted in
Figure \ref{fig_edd} (black and  red histograms) using the methodology
presented in the Appendix.   We estimate a null hypothesis probability
that the two  distributions are drawn from the  same parent population
of 6  and 23 per cent for  AGN in the redshift  intervals 0.1--0.6 and
0.6--1.0, respectively.  Therefore, for  the low redshift subsample we
extended to the Eddington ratio the results of the previous section on
the specific accretion rate.   We find tentative evidence, significant
at  the  $2\sigma$  level,  that  AGN in  star-forming  and  quiescent
galaxies  have  different Eddington  ratio  distributions.  At  higher
redshift, $0.6<z<1.0$,  we find  no evidence for  a difference  in the
accretion  properties  of  AGN  split  by  $UVJ$  colours.   For  that
subsample   however,   if   we   limit   the   comparison   to   $\log
\lambda_{Edd}\ga-3$ we estimate a null hypothesis probability of 2 per
cent  that the  two distributions  are  drawn from  the sample  parent
population.    There  is  therefore   evidence,  significant   at  the
$\approx2\sigma$ level,  that the accretion  properties of AGN  in the
range  $0.6<z<1.0$ and  with $\log  \lambda_{Edd}\ga-3$ depend  on the
level of star-formation of their hosts.

\section{Discussion}

We combine  Chandra data in the  CDFS, AEGIS and  C-COSMOS fields with
UV--to--near-IR  photometry to  place  X-ray AGN  hosts  on the  $UVJ$
diagram  and  split  them  into  quiescent  and  star-forming  systems
independent  of dust induced  biases.  Morphological  evidence further
suggests that grouping  AGN hosts by their $UVJ$  colour selects black
holes that grow their mass under different physical conditions related
to different levels of star-formation rate.

We then estimate the fraction of the accretion density of the Universe
associated with high/low specific star-formation rate AGN hosts in the
redshift range  0.1--1.2.  It is  found that most of  the supermassive
black hole  growth at those  redshifts is associated with  galaxies in
the  high specific star-formation  rate region  of the  $UVJ$ diagram.
Figure \ref{fig_massz}  suggests that this  result may be  a selection
effect of  the optical magnitude  limit ${\cal R}=24$\,mag  applied to
the  X-ray AGN  samples.  Nevertheless,  we  estimate a  total XLF  in
Figure  \ref{fig_xlf} that  agrees well  with the  results of  Aird et
al. (2010).  This indicates that if we miss AGN because of the optical
magnitude limit of the sample, their contribution to the space density
and hence, the integrated X-ray luminosity density, should be small.

Also, our  finding that the accretion  density is dominated  by AGN in
star-forming galaxies is consistent  with studies that link the growth
of SMBHs  to the specific star-formation of  their hosts.  Georgakakis
et al.  (2011) showed that  the evolution with redshift of the optical
and stellar-mass functions of AGN hosts relative to the overall galaxy
population  suggests  that  they  are associated  with  high  specific
star-formation rate systems.  In this picture the rapid decline of the
AGN  space density  at $z<1$  is related  to the  drop of  the average
specific star-formation rate of the overall galaxy population at those
redshifts.   Herschel  data  also  suggest  that AGN  hosts  have,  on
average, specific star-formation rates  similar to or even higher than
galaxies      on       the      main      star-formation      sequence
\citep[e.g.][]{Santini2012,  Rovilos2012}.   Clustering  studies  that
attempt to constrain the distribution of AGN in the cosmic web and not
just their mean dark matter  halos mass, find that a potentially large
fraction of the  population at $z\la1$ lives in  $\rm \log M/M_{\odot}
\approx 12 - 13$ halos \citep{Allevato2012, Mountrichas2013}.  This is
close  to the  characteristic dark  matter halo  mass scale  where the
efficiency of star-formation peaks, in terms of stellar mass over dark
matter  halo   mass  ratio  \citep[e.g.][]{Moster2010,  Leauthaud2012,
  Behroozi2013}.

At the  same time however, we also  find that a fraction  of X-ray AGN
are associated  with early-type hosts  in the quiescent,  low specific
star-formation rate region of the UVJ diagram. This is consistent with
studies that  use the alternative approach of  fitting model templates
to  the observed  spectral  energy distribution  to  account for  dust
reddening \citep{Cardamone2010cmd}.

We  also explore whether  active SMBHs  split by  the $UVJ$  colour of
their hosts  have different accretion properties,  which would suggest
different  fueling modes.   We estimate  the specific  accretion rate,
$\lambda$, of X-ray AGN to approximate the accretion properties of the
central black  hole.  We also convert  $\lambda$ into approximate
Eddington ratios,  $\lambda_{Edd}$, by assuming  a correlation between
black  hole  mass and  bulge  stellar  mass.   For the  low  redshift
sub-sample of Fig.  \ref{fig_edd} we find evidence, significant at the
$2\sigma$ level,  that X-ray AGN  in star-forming and  quiescent hosts
have   different   specific   accretion   rate  or   Eddington   ratio
distributions.  AGN  in star-forming hosts dominate  at high $\lambda$
or $\lambda_{Edd}$, while those in quiescent hosts become increasingly
important toward low Eddington ratios or specific accretion rates.  We
do not find such trends for AGN in the interval $0.6<z<1.0$.  At those
redshifts the  specific accretion rate  distributions of AGN  split by
the level  of star-formation of  their hosts are  consistent. However,
there  is  evidence  significant  at  the  $2\sigma$  level  that  the
Eddington    ratio     distribution    of    the     subsample    with
$\log\lambda_{Edd}\ga-3$ is different  for $UVJ$ selected star-forming
and quiescent  AGN hosts.  Differences at a  similar significance
level in the  specific accretion rate distributions of  obsured AGN at
$z=0.6-4$  split by  the level  of star-formation  of their  hosts was
reported previously by \cite{Brusa2009}. 

The  evidence above  tentatively  suggests the  presence of  different
fuelling  modes  among  the  X-ray  AGN population  at  least  out  to
$z\approx0.6$.  If confirmed with larger samples, this would extend to
higher redshifts results from  local samples ($z\la0.1$), which report
striking differences  at a high statistical significance  level in the
accretion  properties  of SDSS  narrow  optical  emission-line AGN  as
function   of   the   level   of   star-formation   of   their   hosts
\citep{Kauffmann_Heckman2009}.   It is shown  that active  black holes
associated with  the most  actively star-forming galaxies  dominate at
high  Eddington  ratios  and   follow  a  log-normal  distribution  in
$\lambda_{Edd}$.  In contrast, active  SMBHs in quiescent galaxies are
characterised by low Eddington  ratios and a power-law distribution in
$\lambda_{Edd}$.

The  less  pronounced  trends  between  specific  accretion  rate  and
star-formation   in   our   sample   compared  to   the   results   of
\cite{Kauffmann_Heckman2009}  likely  relate  to  differences  in  the
analysis of  the data and ultimately,  observational limitations (e.g.
signal-to-noise ratio,  number statistics) when  performing population
studies  of  AGN  outside  the  local  Universe.   We  use  broad-band
rest-frame colours as  a proxy to star-formation rate  and split X-ray
AGN into two groups,  star-forming and quiescent.  In reality however,
AGN    hosts     span    a    range     of    star-formation    rates.
\cite{Kauffmann_Heckman2009}    find   that   the    Eddington   ratio
distribution of  AGN in their  sample varies smoothly  from log-normal
for  the  subsamples  with  the  highest level  of  star-formation  to
power-law toward the least  star-forming hosts.  This trend is diluted
when splitting  into broad bins of  star-formation. Another difference
between    the     results    presented    here     and    those    of
\cite{Kauffmann_Heckman2009} is the  method adopted to approximate the
mass  of  the  black  hole  and  hence,  determine  the  corresponding
Eddington ratio. We  use the total stellar mass as  proxy of the bulge
mass and therefore estimate the specific accretion rate instead of the
Eddington ratio  of AGN.  We  attempt to correct, at  least in an
approximate way,  for the fact that  the bulge mass of  some AGN hosts
are only a fraction of the total stellar mass. Nevertheless, the lack
of  bulge  mass   proxies  (e.g.   bulge/disk  decomposition,  stellar
velocity  dispersion)  for  individual  sources in  the  sample  could
further dilutes any trends  between Eddington ratio and star-formation
rate.

{\sc galform} is  one of the semi-analytic model  for the cosmological
evolution of  AGN and galaxies  that postulates two channels  of black
hole  growth,   each  of  which  occurs  in   galaxies  with  distinct
star-formation  histories \citep{Bower2006,  Fanidakis2012}.   In that
SAM SMBHs grow when their  hosts experience a starburst event, because
of either  secular processes (e.g.  disk instabilities)  or mergers. A
fraction of the gas that  is available to star-formation is assumed to
accrete onto the SMBH.  Additionally, in {\sc galform} AGN activity is
also triggered  when diffuse hot gas  in quasi-hydrostatic equilibrium
in the parent dark matter halo is accreted onto the SMBH without being
cooled  first onto  the galactic  disk.  The  latter fuelling  mode is
decoupled  from star-formation  and occurs  in passive  galaxies.  The
evolution of  the AGN  population in {\sc  galform} is related  to the
interplay between the two  SMBH fuelling modes.  The starburst channel
is important at high redshift  and high accretion rates, while the hot
halo  accretion mode  dominates  at low  redshifts  and low  accretion
rates.  Moreover, the Eddington ratio distribution in {\sc galform} is
bimodal  as a  result  of the  two  fuelling modes  (see  Figure 1  of
Fanidakis  et al. 2013).   The starburst  mode dominates  at accretion
rates close to Eddington and has  a tail that extends to low Eddington
ratios.  The  hot-halo model becomes important at  low accretion rates
relative to the Eddington limit.  There are therefore similarities, at
least at  the qualitative level,  between the accretion  properties of
AGN in {\sc  galform} and those inferred in the  present paper or from
local samples \citep{Kauffmann_Heckman2009}.

\begin{figure}
\begin{center}
\includegraphics[height=0.9\columnwidth]{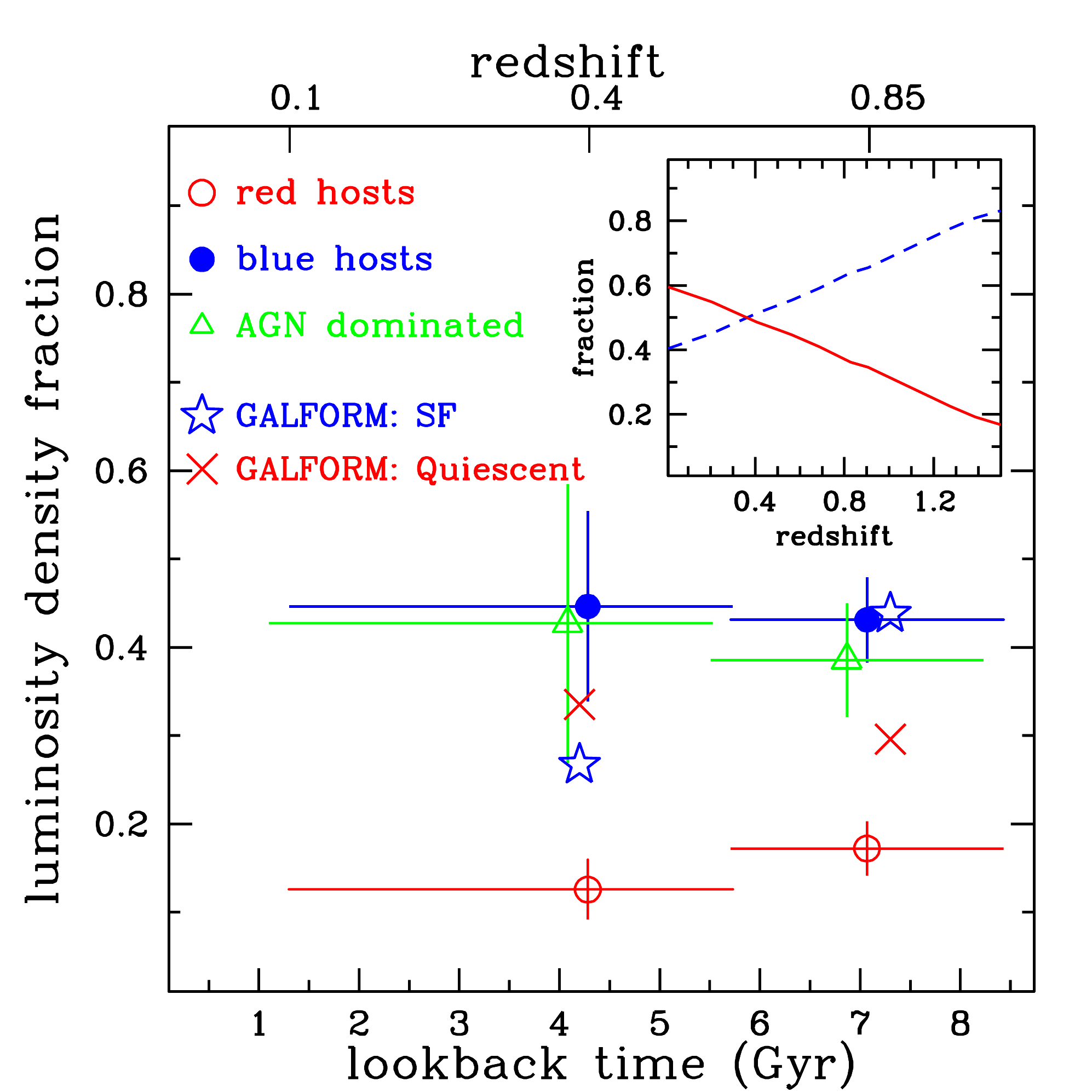}
\end{center}
\caption{The fraction of the  X-ray luminosity density associated with
different  AGN samples  is plotted  as  a function  of look-back  time
(lower x-axis) and redshift (upper  x-axis).  The red open circles are
for red AGN hosts, the blue filled circles represent AGN in blue hosts
and  the  green  triangles  correspond  to AGN  with  optical  colours
contaminated  by  the  central  engine.  The  vertical  errorbars  are
Poisson  estimates  propagated from  the  uncertainties  in the  X-ray
luminosity  density.   The horizontal  errors  represent the  redshift
interval of the different  sub-samples.  For clarity the triangles are
offset by -0.2\,Gyrs.  The inset  plot shows as a function of redshift
the predictions  of {\sc galform} SAM for  the fractional contribution
to the total  X-ray luminosity density of AGN  in $UVJ$-quiescent (red
solid line)  and $UVJ$-star-forming (blue dashed  curve) hosts.  These
results cannot  be directly compared with the  observations because of
the fraction  of broad-line QSOs in  the sample for  which host galaxy
colours  cannot be  determined.   We therefore  correct {\sc  galform}
predictions  at  redshifts  $z=0.4$  and  $z=0.85$  for  the  observed
fraction of  broad-line QSOs in the  sample as described  in the text.
These corrected model predictions are plotted with the blue stars (AGN
hosts in the star-forming region of the $UVJ$ diagram) and red crosses
(AGN   hosts  in  the   quiescent  region   of  the   $UVJ$  diagram).
}\label{fig_ld_fraction}
\end{figure}

\begin{figure}
\begin{center}
\includegraphics[height=0.9\columnwidth]{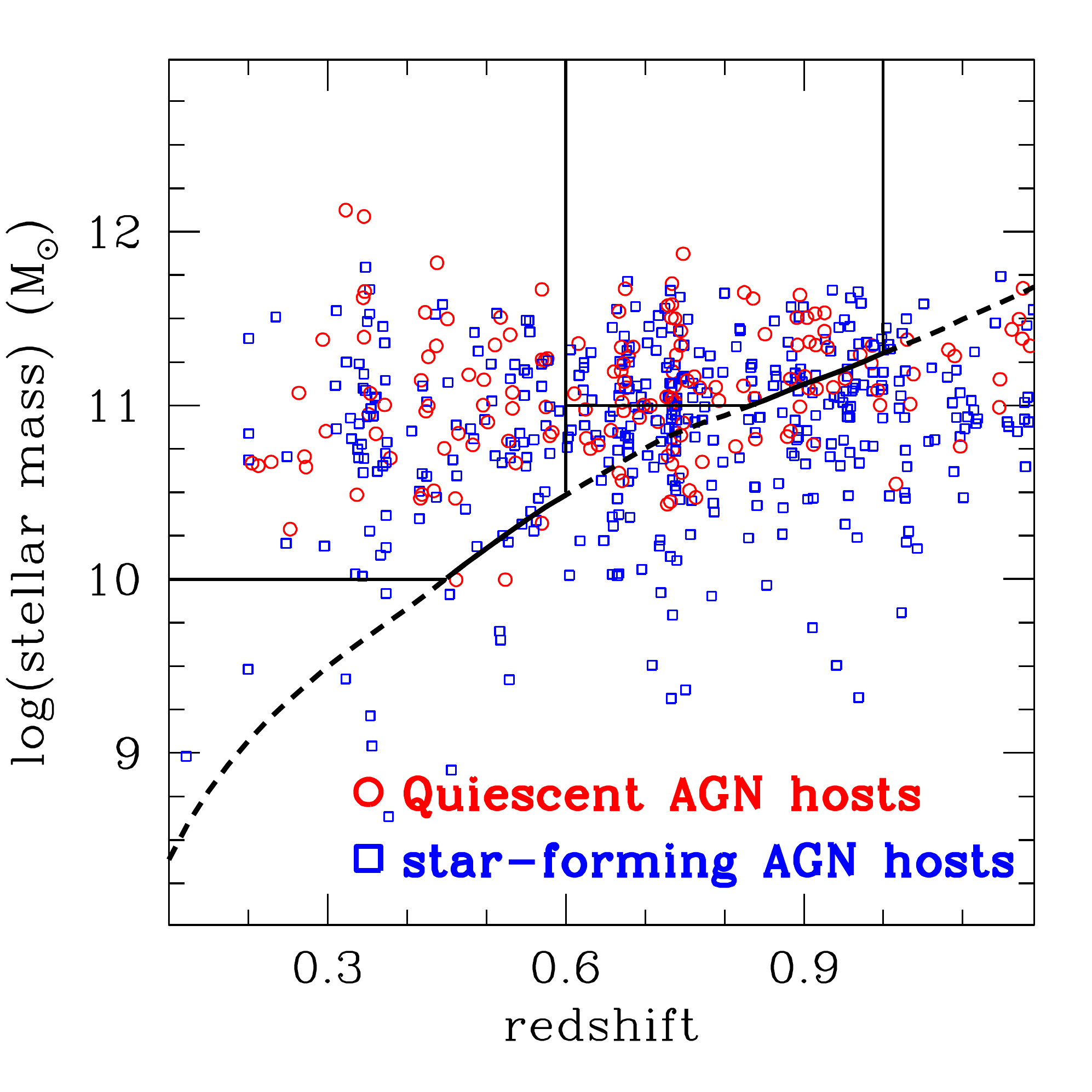}
\end{center}
\caption{X-ray AGN host galaxy stellar mass as a function of redshift.
Red circles  and blue squares are  for AGN hosts in  the quiescent and
star-forming  region of  the $UVJ$  diagram, respectively.   The black
dashed curve  shows the redshift-dependent  mass limit of  a maximally
old  galaxy  with ${\cal  R}=24$\,mag  (see  text  for details).   The
horizontal  solid lines  show the  mass limits  of $10^{10}$  and $\rm
10^{11}\,M_{\odot}$, used to  define nearly volume-limited AGN samples
in  the redshift  intervals $z=0.1-0.6$  and $0.6-1.0$.   The vertical
solid lines mark  the limits of those redshift  intervals.  AGN within
the  wedges defined  by the  continuous sections  of the  black dashed
curve  and  the  vertical  and  horizontal solid  lines  are  used  to
construct specific accretion rate distributions.  }\label{fig_massz}
\end{figure}

\begin{figure*}
\begin{center}
\includegraphics[height=1\columnwidth, angle=270]{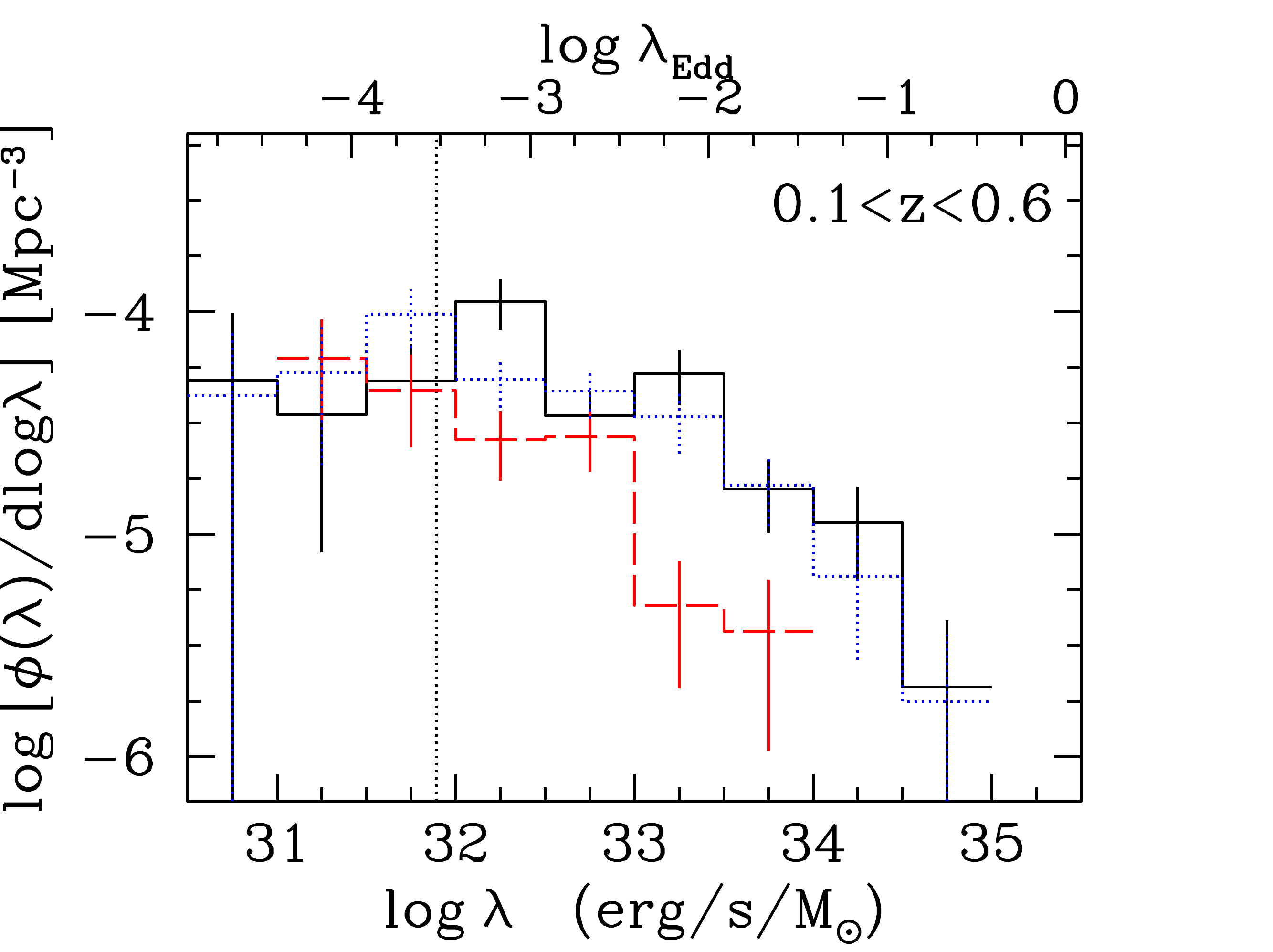}
\includegraphics[height=1\columnwidth, angle=270]{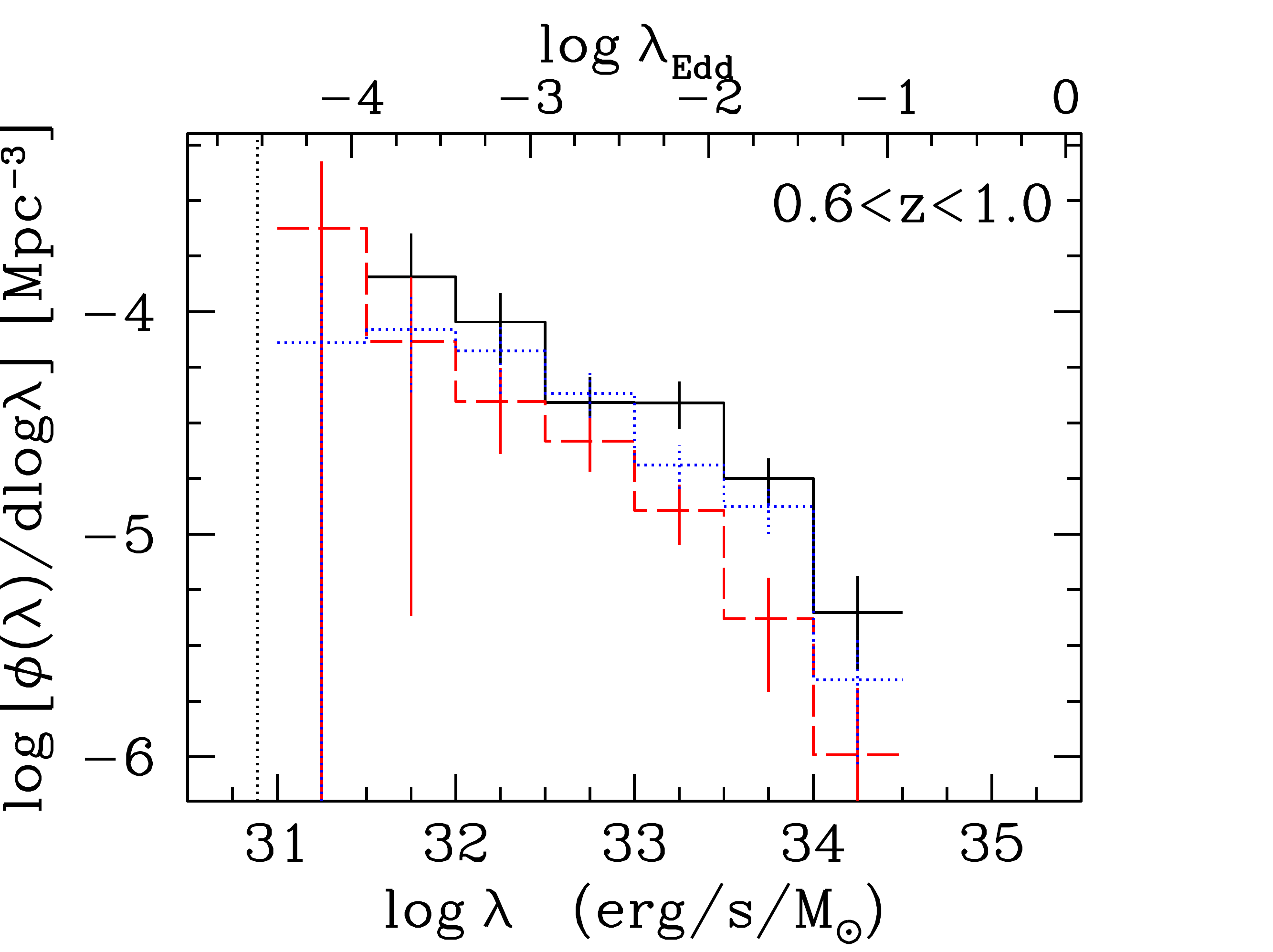}
\end{center}
\caption{Specific accretion rate  distribution, $\lambda$ of X-ray AGN
in  quiescent (red  dashed  histogram) and  star-forming (blue  dotted
histogram) hosts, classified  on the basis of their  UVJ colours.  The
panel on  the left corresponds to  X-ray AGN in  the redshift interval
0.1--0.6. The  panel on  the right  is for X-ray  AGN in  the redshift
range  $z=0.6-1.0$.   The  vertical  dotted line  shows  the  specific
accretion   rate  completeness   limits  for   the  two   samples,  i.e.
$\log\lambda\ga32$  and  $\log\lambda\ga31$  ($\rm erg/s/M_{\odot}$) for the  $z=0.1-0.6$  and
$z=0.6-1.0$ samples respectively.   The  y-axis in  both panels  is
space density in logarithmic bins  of specific accretion rate. The top
y-axis shows  the correspondence  between specific accretion  rate and
Eddington ratio, $\lambda_{Edd}$ under  the assumptions that AGN hosts
are bulge-dominated  (i.e.  bulge mass, $M_{bulge}$,  equals the total
stellar mass of the galaxy) and black hole mass scales with bulge mass
as  $M_{SMBH}=0.002\,M_{bulge}$ \citep{Marconi_Hunt2003}.   The former
assumption  breaks down for  late-type AGN  hosts in  the star-forming
region  of  the  $UVJ$  diagram (e.g.   Fig.   \ref{fig_morph_cosmos},
\ref{fig_morph_aegis}), for  which the bulge  mass is fraction  of the
total   stellar  mass.    The  Eddington   ratio  of   those   AGN  is
underestimated by  a factor equal to  the ratio of the  bulge to total
stellar mass ratio,  $M_{bulge}/M_{total}$.  The black solid histogram
plots  how the  $\lambda_{Edd}$  distribution of  AGN in  star-forming
hosts changes if we assume $M_{bulge}/M_{total}=0.5$ for the late-type
sub-population         and         $M_{bulge}/M_{total}=1$         for
early-types. }\label{fig_edd}
\end{figure*}

A  prediction of {\sc  galform} is  that a  fraction of  the accretion
density of the Universe at a given redshift is associated with passive
low  specific-star-formation  galaxies.   This  is  tested  in  Figure
\ref{fig_ld_fraction}, which compares  the observational data with the
predictions of {\sc galform}. AGN  hosts in that model are first split
into star-forming  and quiescent using the same  $UVJ$ selection wedge
adopted for the real data.  The 2-10\,keV X-ray luminosity function of
the  two  sub-populations  is   then  integrated  to  determine  their
fractional contribution  to the total accretion  density.  The results
are plotted as a function of redshift in Figure \ref{fig_ld_fraction}.
In the comparison of the model with the data we attempt to account for
the fraction  of broad-line AGN  in the observations.   Although there
are  suggestions that such  systems are  mostly found  in star-forming
hosts   at  low   redshift   \citep[$z\la0.1$;][]{Trump2013},  it   is
challenging to  constrain the $UVJ$ colours of  the underlying stellar
population  of the  higher  redshift broad-line  AGN sample  presented
here.  We therefore  choose to correct the model  predictions for this
population.   From the  observations we  estimate at  any  given X-ray
luminosity and  redshift interval  the space density  of the  AGN with
contaminated  colours relative  to the  total XLF,  $f_{QSO}(L_X, z)$.
This fraction is then subtracted from the {\sc galform} model XLF when
integrating to  determine the  starburst and hot-halo  mode luminosity
densities relative to the total.

{\sc  galform} predicts  that the  contributions of  the  hot-halo and
starburst  modes  to  the  accretion  density change  rapidly  and  in
opposite  directions with  redshift,  i.e. see  inset  plot of  Figure
\ref{fig_ld_fraction}.  This trend, although diluted, is still present
even after correcting  the model results for the  observed fraction of
BL   QSOs    in   the   sample    (i.e.    main   panel    of   Figure
\ref{fig_ld_fraction}).  Another  prediction of {\sc  galform} is that
about  30\% of  the  X-ray  luminosity density  at  both $z=0.40$  and
$z=0.85$  is  associated with  AGN  in  quiescent  hosts accreting  in
hot-halo   mode.   These   predictions   are  in   tension  with   the
observational  results plotted  in  Figure \ref{fig_ld_fraction}.   We
find that only  up to 20\% of the X-ray  luminosity density at $z\la1$
is  associated with AGN  hosts in  the quiescent  region of  the $UVJ$
diagram. Also  the fraction of  the accretion density  associated with
star-forming/quiescent hosts  does not  change within the  errors from
$z=0.85$ to $z=0.40$.  One  possible solution to these discrepancies is
to  relax the  tight correspondence  between accretion  mode  and host
galaxy $UVJ$ colours in {\sc galform}. Hot-halo mode AGN hosts in that
model do not  form new stars and are  therefore found predominantly in
the quiescent part of the  $UVJ$ diagram.  In contrast, starburst mode
AGN hosts  populate almost exclusively the star-forming  region of the
$UVJ$ diagram.  The assumption of the model that hot-halo mode AGN are
completely  disjoint  from  star-formation is  probably  conservative.
Some level of  star-formation might be expected in  the hosts of those
AGN  as  the  hot  gas  cools  from galactic  scales  onto  the  black
hole. Allowing for this effect  could shift a fraction of the hot-halo
AGN in  {\sc galform} into  the blue region  of the $UVJ$  diagram. In
this  respect   it  is  important   that  {\sc  galform}   predicts  a
sufficiently large pool of hot-halo  AGN, which could populate the red
region of the UVJ diagram.

\section{Conclusions}

We  use  the  $UVJ$  diagram  to split  AGN  hosts  into  star-forming
(including dust reddened) and quiescent. The host-galaxy morphology of
the  two sub-populations  is found  to be  different,  suggesting that
selection on the  $UVJ$ diagram provides a means  of identifying SMBHs
that grow their mass under  distinct physical conditions. AGN hosts in
the  quiescent  region  of  the  $UVJ$ diagram  are  early-type  bulge
dominated  galaxies.  In  contrast, star-forming  AGN hosts  include a
large fraction  (about 50\%) of  late-type systems.  We  also estimate
the  fraction  of the  accretion  density  associated  with those  two
classes of  hosts at  redshifts $z\approx0.40$ and  0.85. Most  of the
accretion density  at those redshifts is taking  place in star-forming
hosts. Nevertheless,  about 15-20\% of  the AGN luminosity  density is
associated with  galaxies in  the quiescent part  of the  UVJ diagram.
There is also  evidence, significant at the $2\sigma$  level, that AGN
in the low redshift  subsample ($0.1<z<0.6$) have accretion properties
that depend  on the  level of star-formation  of their hosts.   AGN in
star-forming hosts  dominate at high Eddington ratios,  while those in
quiescent  hosts become  increasingly important  toward  low Eddington
ratios.  At higher redshift, $0.6<z<1.0$, such differences are present
at  the $2\sigma$  level  only  for AGN  with  Eddington ratios  $\log
\lambda_{Edd}\ga-3$.  These results are  consistent with two modes for
growing  black  holes  that  take  place in  galaxies  with  different
star-formation  properties.  We  compare those  observations  with the
predictions  of {\sc  galform} SAM,  which postulates  two  black hole
growth channels, one linked  to star-formation and the other occurring
in passive systems.  This SAM predicts a fraction of accretion density
in quiescent  hosts that  is larger than  the observed.  We  also find
that the  evolution with redshift  of the X-ray luminosity  density of
hot-halo/starburst mode AGN is inconsistent with the observations.  We
argue that  these discrepancies could be attributed  to the assumption
of {\sc galform}  that the hot-halo accretion mode  is not accompanied
with some level  of star-formation in the host  galaxy.  Relaxing this
requirement could bring the model in better agreement with the data.

\section{Acknowledgments}

The  authors  wish to  thank  the  referee,  M. Brusa,  for  providing
constructive comments and suggestions. PGP-G acknowledges support from
the Spanish  Programa Nacional  de Astronom\'ia y  Astrof\'isica under
grant  AYA2012-31277.    This  work  has  made  use   of  the  Rainbow
Cosmological Surveys  Database, which  is operated by  the Universidad
Complutense  de  Madrid  (UCM),   partnered  with  the  University  of
California Observatories  at Santa Cruz  (UCO/Lick,UCSC).  Funding for
the  DEEP2 Galaxy Redshift  Survey has  been provided  in part  by NSF
grants   AST95-09298,  AST-0071048,   AST-0071198,   AST-0507428,  and
AST-0507483 as  well as NASA  LTSA grant NNG04GC89G.  Funding  for the
DEEP3  Galaxy  Redshift  Survey   has  been  provided  by  NSF  grants
AST-0808133,  AST-0807630, and AST-0806732.  This work  benefited from
the {\sc thales} project 383549 that is jointly funded by the European
Union  and the  Greek Government  in  the framework  of the  programme
``Education and lifelong learning''.

\bibliography{/home/age//soft9/BIBTEX/mybib}{}
\bibliographystyle{mn2e}

\appendix
\section{Statistical comparison of accretion rate distribution of AGN
samples}

This section describes the  methodology followed to assess differences
in the  specific accretion rate  distribution of AGN in  quiescent and
star-forming  hosts.  We  use  the Kolmogorov--Smirnov  (K-S)  test  to
estimate  the  probability  of   the  null  hypothesis  that  the  two
distributions are drawn from the same parent population. 

The  K-S test cannot  be used  to compare  directly the  space density
distribution of AGN in specific  accretion rate bins plotted in Figure
\ref{fig_edd}. This is because  those distributions are constructed by
normalising  each  source in  the  sample  by  $V_{max}$ (see  section
\ref{sec_xlf}), i.e.  correcting for the selection function.  Also, it
is not  possible to  apply the K-S  test to the  ``observed'' specific
accretion rate  distributions, i.e.   those constructed by  summing up
AGN without applying any $V_{max}$ corrections.  This is because small
differences  in  the  selection  functions  of AGN  in  quiescent  and
star-forming  hosts, e.g.  X-ray  luminosity distribution,  could bias
any results.

The  approach we  follow instead  starts with  a model  for  the space
density of AGN as a function of specific accretion rate for one of the
two samples we wish to compare, for example AGN in star-forming hosts.
This  is then  convolved with  the  selection function  of the  second
sample, i.e.  in  this example AGN in quiescent  hosts.  The resulting
model distribution  can then be compared to  the ``observed'' specific
accretion rate  distribution of AGN  in quiescent hosts using  the K-S
test.  The  underlying assumption is  that the two  samples, quiescent
and star-forming, are drawn from the same parent population.

\begin{figure}
\begin{center}
\includegraphics[height=0.9\columnwidth, angle=0]{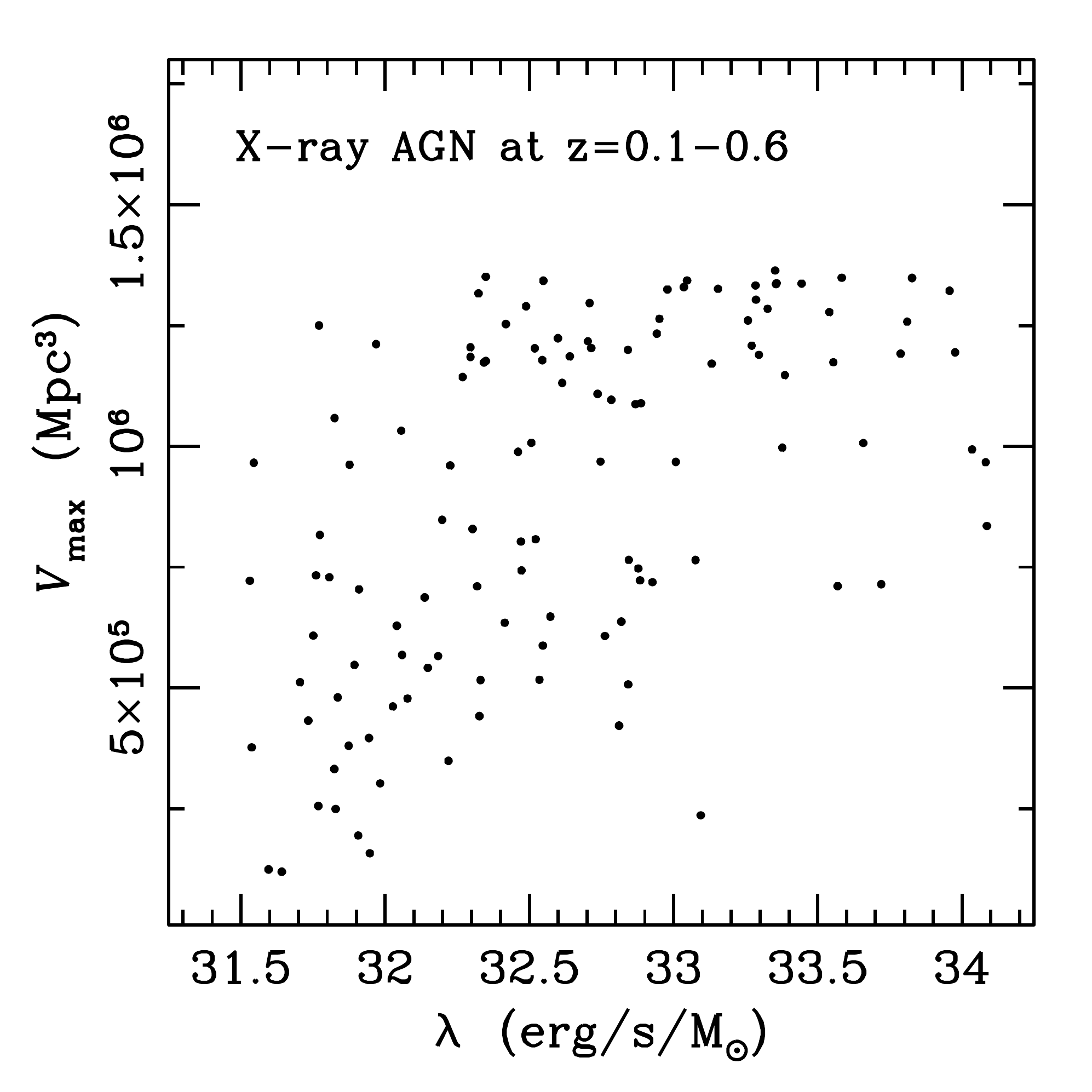}
\end{center}
\caption{Maximum volume, $V_{max}$ as a function of specific accretion
rate,  $\lambda$, of  X-ray  AGN in  the  redshift interval  0.1--0.6.
}\label{fig_app_vmax}
\end{figure}

We  choose as model  for the  space density  of AGN  as a  function of
accretion  rate, $\phi(\lambda)$, the  one inferred  from observations
(i.e.  Figure \ref{fig_edd}).  The  selection function of each sample,
quiescent  or   star-forming,  is  essentially   encapsulated  in  the
$V_{max}$   estimated   for   individual   sources.   To   the   first
approximation we  can therefore use  those discreet values  to account
for selection effects.  Figure \ref{fig_app_vmax} shows that $V_{max}$
is  a function of  specific accretion  rate, in  the sense  that lower
specific accretion  rate systems have, on  average, smaller $V_{max}$,
i.e.  they drop from the  sample at lower redshift, compared to higher
accretion rate systems.

The  convolution  of   $\phi(\lambda)$  with  the  selection  function
proceeds as follows.  First  we sample from $\phi(\lambda)$ to produce
a series of  $\lambda$ values. The AGN from  the real sample catalogue
with specific accretion  rate closest to each of  the random $\lambda$
draws  is  identified. The  $V_{max}$  corresponding  to  that AGN  is
assigned to  the $\lambda$  randomly drawn from  $\phi(\lambda)$. This
$V_{max}$  value  is  then  used  to determine  if  the  corresponding
$\lambda$ should  be retained in the  sample.  A random  number in the
range   $0-1$   is  produced   and   is   compared   with  the   ratio
$V_{max}/max(V_{max})$,  where $max(V_{max})$  is the  maximum  of all
$V_{max}$  in  the  sample.   If   the  random  number  is  less  than
$V_{max}/max(V_{max})$  then  $\lambda$   is  kept,  otherwise  it  is
discarded.  These  steps are repeated  for all $\lambda$  values drawn
from  the model  $\phi(\lambda)$.  The  retained $\lambda$  values are
used to build the  cumulative probability distribution function of the
model. The  K-S test  is then applied  to compare the  distribution of
specific accretion  rates drawn from  the model after  convolving with
the  selection  function  with  that inferred  from  the  observations
without applying any $V_{max}$ corrections to individual sources.

\begin{figure}
\begin{center}
\includegraphics[height=0.9\columnwidth, angle=0]{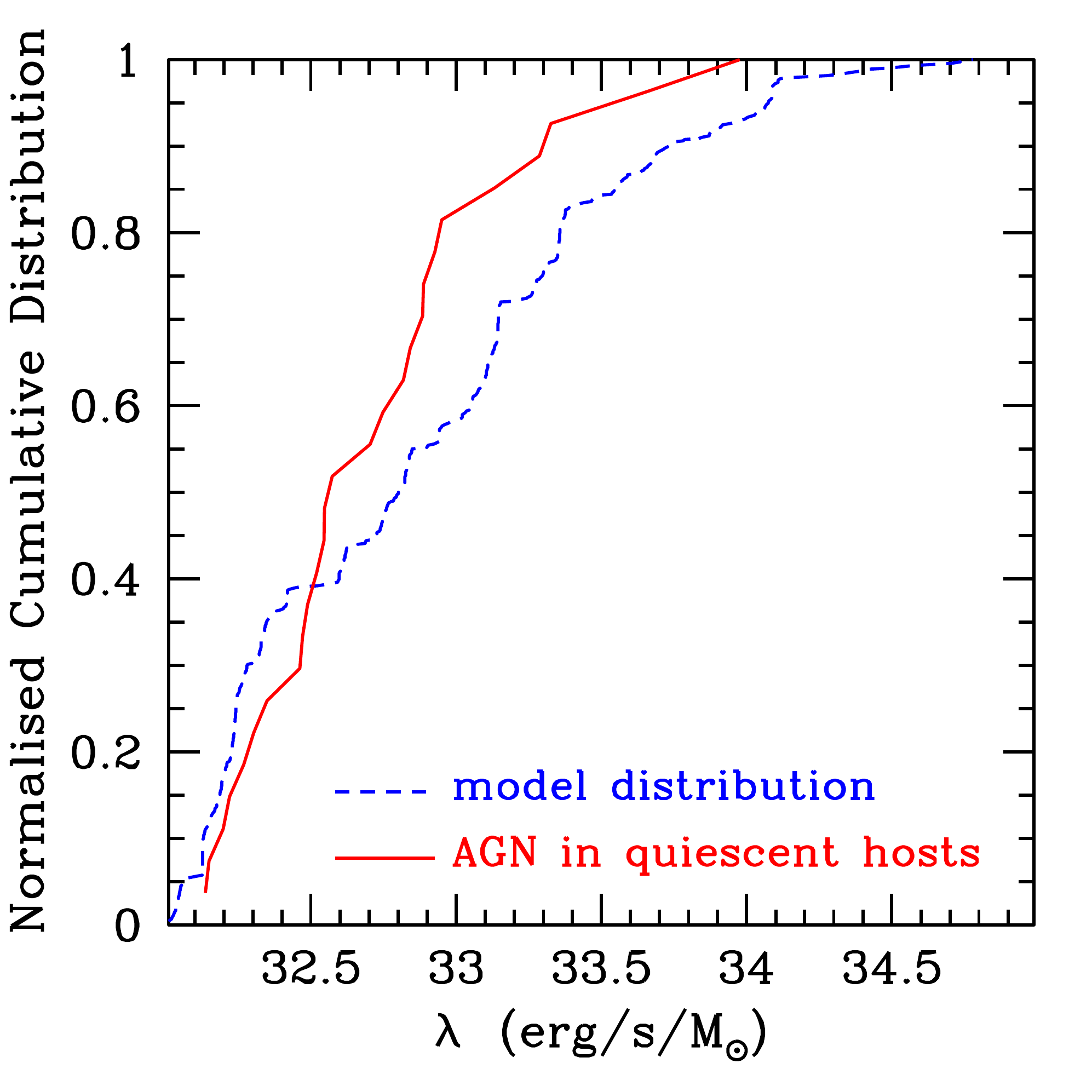}
\end{center}
\caption{Normalised cumulative  distribution of  AGN as a  function of
specific accretion rate, $\lambda$.  The red solid curve is for AGN in
quiescent hosts in the  reshift interval 0.1--0.6. This is constructed
by  summing  up  sources   without  applying  any  selection  function
corrections.  The  blue dashed  distribution is the  comparison sample
constructed from  the space density  of AGN in star-forming  hosts and
convolving with the  selection function of the AGN  in quiescent hosts
(see  text  for  details).  The  K-S  test  can  be applied  to  those
distributions to estimate the null hypothesis that the two populations
are drawn from the same parent population.  }\label{fig_app_cum}
\end{figure}

As an example, we compare  the specific accretion rate distribution of
AGN   in  quiescent/star-forming  hosts   in  the   redshift  interval
$z=0.1-0.6$ plotted  in \ref{fig_edd}-left.  We  only consider sources
with  $\log\lambda>32$  ($\rm  erg/s/M_{odot}$).   At  lower  specific
accretion  rates  incompleteness  is  affecting  the  estimated  space
density of AGN.  We use  as model the inferred $\phi(\lambda)$ for AGN
in star-forming  hosts plotted  in Fig.  \ref{fig_edd}-left.   This is
then convolved with the selection  function of the quiescent AGN host.
The resulting cumulative distribution for AGN in star-forming hosts is
plotted in  Figure \ref{fig_app_cum}.  Also plotted is  that figure is
the   ``observed''  cumulative  distribution   of  AGN   in  quiescent
hosts. The K-S  test is used to estimate the  probability that the two
distributions  are  drawn  from  same  parent  population.   The  null
hypothesis  has  a probability  of  5  per  cent.  Therefore,  in  the
redshift interval $z=0.1-0.6$ AGN in quiescent/star-forming hosts have
different  specific accretion rate  distributions at  the 95  per cent
confidence  level,  or about  $2.0\sigma$  in  the  case of  a  Normal
distribution.

\end{document}